\pdfoutput=1
\documentclass[a4paper,12pt]{article}
\usepackage{jheppub}
\usepackage{amsmath}
\usepackage{amssymb}
\usepackage{braket}
\usepackage{cancel}
\usepackage{graphicx}
\usepackage[utf8]{inputenc}
\usepackage{hyperref}
\usepackage{multirow}
\usepackage{slashed}
\usepackage{booktabs}
\usepackage[normalem]{ulem}
\usepackage{wasysym}
\usepackage{braket}
\usepackage{simplewick}
\usepackage{array}
\usepackage{float}
\usepackage{titlesec}
\usepackage{listings}
\usepackage{bbold}
\usepackage[english]{babel}
\usepackage{subcaption}
\usepackage{braket}
\usepackage{comment}
\usepackage{enumitem}
\usepackage{bm}
\usepackage{verbatim}
\usepackage{scalefnt}
\usepackage[export]{adjustbox}
\allowdisplaybreaks

\newcommand{\re}{\mathrm{e}}

\newcommand{\alphas}{\alpha_{\mathrm{s}}}

\newcommand{\CF}{C_\mathrm{F}}
\newcommand{\CA}{C_\mathrm{A}}
\newcommand{\dF}{d_\mathrm{F}}
\newcommand{\dA}{d_\mathrm{A}}
\newcommand{\TF}{T_\mathrm{F}}
\newcommand{\Nc}{N_\mathrm{c}}
\newcommand{\nf}{n_\mathrm{f}}
\newcommand{\nl}{n_\mathrm{l}}
\newcommand{\nh}{n_\mathrm{h}}
\newcommand{\nm}{n_\mathrm{m}}

\newcommand{\zmFFFF}{z_m^\mathrm{FFFF}}
\newcommand{\zmFFFA}{z_m^\mathrm{FFFA}}
\newcommand{\zmFFAA}{z_m^\mathrm{FFAA}}
\newcommand{\zmFAAA}{z_m^\mathrm{FAAA}}
\newcommand{\zmdFA}{z_m^{d_\mathrm{FA}}}
\newcommand{\zmdFFL}{z_m^{d_\mathrm{FF}\mathrm{L}}}
\newcommand{\zmdFFH}{z_m^{d_\mathrm{FF}\mathrm{H}}}
\newcommand{\zmFFFL}{z_m^\mathrm{FFFL}}
\newcommand{\zmFFAL}{z_m^\mathrm{FFAL}}
\newcommand{\zmFAAL}{z_m^\mathrm{FAAL}}
\newcommand{\zmFFLL}{z_m^\mathrm{FFLL}}
\newcommand{\zmFALL}{z_m^\mathrm{FALL}}
\newcommand{\zmFLLL}{z_m^\mathrm{FLLL}}
\newcommand{\zmFFFH}{z_m^\mathrm{FFFH}}
\newcommand{\zmFFAH}{z_m^\mathrm{FFAH}}
\newcommand{\zmFAAH}{z_m^\mathrm{FAAH}}
\newcommand{\zmFFHH}{z_m^\mathrm{FFHH}}
\newcommand{\zmFAHH}{z_m^\mathrm{FAHH}}
\newcommand{\zmFHHH}{z_m^\mathrm{FHHH}}
\newcommand{\zmFFLH}{z_m^\mathrm{FFLH}}
\newcommand{\zmFALH}{z_m^\mathrm{FALH}}
\newcommand{\zmFLLH}{z_m^\mathrm{FLLH}}
\newcommand{\zmFLHH}{z_m^\mathrm{FLHH}}
\newcommand{\zmFMLL}{z_m^\mathrm{FMLL}}
\newcommand{\zmFMLH}{z_m^\mathrm{FMLH}}
\newcommand{\zmFMHH}{z_m^\mathrm{FMHH}}
\newcommand{\zmFMML}{z_m^\mathrm{FMML}}
\newcommand{\zmFMMH}{z_m^\mathrm{FMMH}}
\newcommand{\zmFMMM}{z_m^\mathrm{FMMM}}
\newcommand{\zmFFML}{z_m^\mathrm{FFML}}
\newcommand{\zmFAML}{z_m^\mathrm{FAML}}
\newcommand{\zmFFMH}{z_m^\mathrm{FFMH}}
\newcommand{\zmFAMH}{z_m^\mathrm{FAMH}}
\newcommand{\zmFFMM}{z_m^\mathrm{FFMM}}
\newcommand{\zmFAMM}{z_m^\mathrm{FAMM}}
\newcommand{\zmFFFM}{z_m^\mathrm{FFFM}}
\newcommand{\zmFFAM}{z_m^\mathrm{FFAM}}
\newcommand{\zmFAAM}{z_m^\mathrm{FAAM}}
\newcommand{\zmdFFM}{z_m^{d_\mathrm{FF}\mathrm{M}}}

\newcommand{\zmM}{z_m^\mathrm{M}}
\newcommand{\zmMfour}{z_m^{\mathrm{M},(4)}}
\newcommand{\ymM}{y_m^\mathrm{M}}

\usepackage{lscape}
\usepackage[dvipsnames]{xcolor}

\begin{document}

\preprint{
\begin{flushright}
CERN-TH-2026-268,
P3H-26-087,
TTP26-040,
ZU-TH 46/26
\end{flushright}
}

\title{
\boldmath
Four-loop $\overline{\rm MS}$-pole quark mass relation with two different masses
}

\author[a,b]{Matteo Fael,}
\author[c,d]{Fabian Lange,}
\author[e]{Kay Sch\"onwald,}
\author[f]{and Matthias Steinhauser}

\affiliation[a]{
Dipartimento di Fisica e Astronomia ``G.~Galilei,'' Università di Padova,
\\
Via F.~Marzolo 8, 35131
Padova, Italy
}

\affiliation[b]{
Istituto Nazionale di Fisica Nucleare, Sezione di Padova,
\\ Via F.~Marzolo 8, 35131 Padova, Italy
}

\affiliation[c]{
Physik-Institut, Universität Zürich,\\
Winterthurerstrasse 190, 8057 Zürich, Switzerland}

\affiliation[d]{PSI Center for Neutron and Muon Sciences,\\
5232 Villigen PSI, Switzerland}

\affiliation[e]{
Theoretical Physics Department, CERN,\\
1211 Geneva, Switzerland}

\affiliation[f]{
Institut f\"ur Theoretische Teilchenphysik, Karlsruhe Institute of Technology (KIT),\\
Wolfgang-Gaede-Stra\ss{}e 1,
76131 Karlsruhe, Germany}

\emailAdd{matteo.fael@pd.infn.it}
\emailAdd{fabian.lange@physik.uzh.ch}
\emailAdd{kay.schonwald@cern.ch}
\emailAdd{matthias.steinhauser@kit.edu}

\abstract{

We calculate the relation between heavy-quark masses defined in
the on-shell and $\overline{\rm MS}$ renormalization schemes up to
four-loop order in Quantum Chromodynamics, accounting for a second,
non-zero quark mass. We employ a semi-analytical method that yields
precise results valid for all values of the mass ratio. In particular,
our approach enables generalized expansions for small and large values
of the second quark mass, incorporating all logarithmic
contributions. For the limiting cases of a vanishing or equal second
quark mass, we significantly improve upon the precision of known results for 18 of the
23 colour structures.

}

\maketitle
\flushbottom

\newpage


\section{Introduction}

The relation between the pole (or on-shell) and the heavy-quark mass
in the $\overline{\rm MS}$ scheme is a crucial component of many
calculations within perturbative Quantum Chromodynamics (QCD). The
pole mass is based on an intuitive definition linked to the position
of the pole in the perturbative propagator. In contrast, the
$\overline{\rm MS}$ mass is a quantity defined at short distances that
often exhibits improved perturbative behaviour. Since numerous
phenomenological quantities depend on precise relations between pole
and $\overline{\rm MS}$ masses, it is essential to know this
relationship with high accuracy.

The one-loop correction relating the pole mass to the $\overline{\rm
MS}$ mass was calculated more than 40 years ago~\cite{Tarrach:1980up}, and
two-loop corrections have been known since the early
1990s~\cite{Gray:1990yh}.
At two-loop order, contributions involving a closed quark loop appear for the first time.
The mass of the virtual quarks can be zero, equal to the mass of the external quark, or take on another value.
The massless approximation is well justified in many applications
but becomes insufficient whenever the effects of a second heavy quark
cannot be neglected. A prominent example is the bottom-quark mass,
where virtual charm-quark contributions produce corrections that are
numerically important.

The three-loop relation was first calculated using a semi-numerical
approach in Refs.~\cite{Chetyrkin:1999ys,Chetyrkin:1999qi}. This result was confirmed by
analytical calculations in Refs.~\cite{Melnikov:2000qh,Marquard:2007uj}. Contributions
involving two mass scales were calculated a few years later
in Ref.~\cite{Bekavac:2007tk}, with some components determined numerically. Fully
analytical three-loop results for the two-mass contributions are
available from Ref.~\cite{Fael:2020bgs}. Furthermore, this paper presents
detailed expansions for the limiting cases of small, large, and equal
masses, which often prove useful in practical applications.
In Ref.~\cite{Fael:2020bgs} also terms of order $D-4$ have been computed at three
loops which are an important ingredient at four loops.

To date, the result for the single-mass configuration at the four-loop
level has been reported by only one group~\cite{Marquard:2015qpa,Marquard:2016dcn}, with a large
portion of the master integrals calculated numerically. Consequently,
only one or two significant digits could be determined for some of the
individual colour structures. For the final physical results --
obtained after substituting the numerical values for the QCD colour
factors -- two to three significant digits were achieved, which is
sufficient for the current phenomenological applications.

In this paper, we investigate the relation between the
$\overline{\rm MS}$ and on-shell quark masses, taking into account two
heavy quark flavours. We retain the exact dependence on the ratio of
the two masses and derive high-precision numerical results for the
corresponding renormalization constant. Particular attention is paid
to phenomenologically relevant mass hierarchies. In particular, we
provide expansions for the limiting cases of small, large, and equal masses.
Besides the newly computed contributions of a second heavy quark, our results also provide an independent calculation of all fermionic contributions of Refs.~\cite{Marquard:2015qpa,Marquard:2016dcn}, significantly increasing the precision of the corresponding coefficients.

The presence of a second, non-zero mass scale significantly increases
the complexity of multi-loop calculations. Feynman integrals depend on
the mass ratio in a non-trivial way. We describe in detail the
semi-analytical method we developed~\cite{Fael:2021kyg}, which allows for high-precision
results to be obtained in such cases across the entire valid range of
the mass ratio.

The remainder of this paper is organized as follows: In
Section~\ref{sec::notation} we introduce the notation, and in
Section~\ref{sec::techniques} we describe the methods for
calculating the four-loop Feynman integrals. Section~\ref{sec::ren}
deals with renormalization, and our results regarding the relation for
the quark masses are discussed in Section~\ref{sec::results}.
Section~\ref{sec::concl} concludes the paper.

\section{\label{sec::notation}Notation}

In our calculation, we consider two heavy quarks with masses $m_1$ and
$m_2$ and introduce $\nh=1$ and $\nm=1$ to account for the number of
corresponding closed fermion loops. Furthermore, we include $\nl$
massless quarks, such that the number of flavours is given by
$\nf=\nh+\nm+\nl$.
It should be noted that the ratio $m_2/m_1$ can take any value between
$0$ and $\infty$.

We compute the renormalization constant of $m_1$, $Z_m^{\rm OS}$, and
construct the finite quantity \begin{eqnarray} z_m &=& \frac{Z_m^{\rm
OS}}{Z_m^{\overline{\rm MS}}} \,\,=\,\, \frac{m_1^{\overline{\rm MS}}}
{m_1^{\rm OS}} \,.  \end{eqnarray}

For the perturbative expansion of $z_m$ we introduce
\begin{eqnarray}
  z_m =  1 + \sum_{i\ge1} z_m^{(i)} \left(\frac{\alphas(\mu)}{\pi}\right)^i \,,
  \label{eq::zmexp}
\end{eqnarray}
where $\alphas(\mu)$ is the strong coupling constant defined
in the $(\nh+\nm+\nl)$-flavour theory.
We use analogous expansions for all other quantities.

For a generic SU$(\Nc)$ group
the four-loop term of Eq.~(\ref{eq::zmexp}) can be
decomposed into 39 colour structures which are given by
\begin{eqnarray}
  z_m^{(4)} &=&
  \CF^4 \zmFFFF
          +  \CF^3 \CA \zmFFFA
          +  \CF^2 \CA^2 \zmFFAA
          +  \CF \CA^3 \zmFAAA
\nonumber\\&&\mbox{}
          +  \frac{\dF^{abcd}\dA^{abcd}}{\Nc} \zmdFA
          +  \nl \frac{\dF^{abcd}\dF^{abcd}}{\Nc} \zmdFFL
          +  \nh \frac{\dF^{abcd}\dF^{abcd}}{\Nc} \zmdFFH
\nonumber\\&&\mbox{}
          +  \CF^3 \TF \nl \zmFFFL
          +  \CF^2 \CA \TF \nl \zmFFAL
          +  \CF \CA^2 \TF \nl \zmFAAL
\nonumber\\&&\mbox{}
          +  \CF^2 \TF^2 \nl^2 \zmFFLL
          +  \CF \CA \TF^2 \nl^2 \zmFALL
          +  \CF \TF^3 \nl^3 \zmFLLL
\nonumber\\&&\mbox{}
          +  \CF^3 \TF \nh \zmFFFH
          +  \CF^2 \CA \TF \nh \zmFFAH
          +  \CF \CA^2 \TF \nh \zmFAAH
\nonumber\\&&\mbox{}
          +  \CF^2 \TF^2 \nh^2 \zmFFHH
          +  \CF \CA \TF^2 \nh^2 \zmFAHH
          +  \CF \TF^3 \nh^3 \zmFHHH
\nonumber\\&&\mbox{}
          +  \CF^2 \TF^2 \nl \nh \zmFFLH
          +  \CF \CA \TF^2 \nl \nh \zmFALH
          +  \CF \TF^3 \nl^2 \nh \zmFLLH
\nonumber\\&&\mbox{}
          +  \CF \TF^3 \nl \nh^2 \zmFLHH
\nonumber\\&&\mbox{}
        + \CF \TF^3 \nm \nl^2   \zmFMLL
        + \CF \TF^3 \nm \nl \nh \zmFMLH
        + \CF \TF^3 \nm \nh^2   \zmFMHH
          \nonumber\\&&\mbox{}
        + \CF \TF^3 \nm^2 \nl   \zmFMML
        + \CF \TF^3 \nm^2 \nh   \zmFMMH
        + \CF \TF^3 \nm^3       \zmFMMM
          \nonumber\\&&\mbox{}
        + \CF^2 \TF^2 \nm \nl   \zmFFML
        + \CF \CA \TF^2 \nm \nl  \zmFAML
        \nonumber\\&&\mbox{}
        + \CF^2 \TF^2 \nm \nh   \zmFFMH
        + \CF \CA \TF^2 \nm \nh  \zmFAMH
        + \CF^2 \TF^2 \nm^2   \zmFFMM
          \nonumber\\&&\mbox{}
          + \CF \CA \TF^2 \nm^2  \zmFAMM
          +  \CF^3 \TF \nm \zmFFFM
          +  \CF^2 \CA \TF \nm \zmFFAM
\nonumber\\&&\mbox{}
    +  \CF \CA^2 \TF \nm \zmFAAM
    +  \nm \frac{\dF^{abcd}\dF^{abcd}}{\Nc} \zmdFFM
          \,.
          \label{eq::zm4_col}
\end{eqnarray}
The first 23 colour structures are already present for $\nm=0$; the
remaining 16 colour structures shown in the last six lines of Eq.~(\ref{eq::zm4_col})
are proportional to $\nm^3$, $\nm^2$ or $\nm$ and are the subject to this paper.
Analytic and numerical results for the eight coefficients
$\zmFMLL$, $\zmFMLH$, $\zmFMHH$, $\zmFMML$, $\zmFMMH$, $\zmFMMM$,
$\zmFFML$ and $\zmFAML$ can be found in Ref.~\cite{Fael:2021kyg}. Results for the remaining
eight $\nm$ structures are presented here for the first time.

In Eq.~(\ref{eq::zm4_col})
$\CF$ and $\CA$ are the eigenvalues of the quadratic Casimir operators
of the fundamental and adjoint representation for the SU$(\Nc)$ colour group,
respectively. $\TF=1/2$ is the index of the fundamental representation.
$\dF^{abcd}$ and $\dA^{abcd}$ are the symmetrized traces of four generators in
the fundamental and adjoint representation, respectively.  The colour
structures in Eq.~(\ref{eq::zm4_col}) are related to $\Nc$ via
\begin{eqnarray}
  &&\CF = \frac{\Nc^2-1}{2\Nc}\,,\qquad \CA = \Nc\,,
  \nonumber\\
  &&\dF^{abcd}\dF^{abcd} = \frac{(\Nc^2-1)(\Nc^4 - 6 \Nc^2 + 18)}{96 \Nc^2}\,,
  \nonumber\\
  &&\dF^{abcd}\dA^{abcd} = \frac{\Nc (\Nc^2-1) (\Nc^2+6)}{48}\,.
\end{eqnarray}
In the case of QCD we have $\Nc=3$ which leads to
\begin{eqnarray}
  &&\CF = \frac{4}{3}\,,\qquad \CA = 3\,,
  \nonumber\\
  &&\dF^{abcd}\dF^{abcd} = \frac{5}{12}
  \,,\qquad
  \dF^{abcd}\dA^{abcd} = \frac{15}{2} \,.
\end{eqnarray}

It is convenient to introduce the $\nm$-dependent
contribution for $z_m$ as
\begin{eqnarray}
          \zmM &=& z_m - z_m(\nm=0)\,.
\end{eqnarray}
For the corresponding inverse functions we introduce
\begin{align}
    y_m &= \frac{1}{z_m}\,, &
    \ymM &=
    y_m - y_m(\nm=0)\,.
\end{align}

Depending on the application, it is advantageous to transform either
$m_1$, $m_2$, or both into the $\overline{\rm MS}$ scheme. For this
purpose, it is convenient to introduce the variables
\begin{eqnarray}
  x = \frac{m_2^{\rm OS}}{m_1^{\rm OS}}\,,\quad
  x_f(\mu_f) = \frac{\overline{m}_2(\mu_f)}{m_1^{\rm OS}}\,,\quad
  x_q(\mu) = \frac{m_2^{\rm OS}}{\overline{m}_1(\mu)}\,,\quad
  x_{fq}(\mu_f,\mu) = \frac{\overline{m}_2(\mu_f)}{\overline{m}_1(\mu)}\,,
  \label{eq::x_fq}
\end{eqnarray}
where $\mu_f$ is the renormalization scale of the quark mass $m_2$ and $\mu$
is the common renormalization scale of $m_1$ and $\alphas$.

Our starting point is the quantity $z_m \equiv z_m(x)$. We use
$z_m(x_q)$ when $m_1^{\rm OS}$ is transformed into the $\overline{\rm
MS}$ scheme and an expansion in $\alphas$ is subsequently
performed. Analogous notation is used for other scheme transformations
in accordance with Eq.~(\ref{eq::x_fq}).

\section{\label{sec::techniques}Technicalities}

\begin{figure}[t]
\centering
\includegraphics[width=0.24\textwidth]{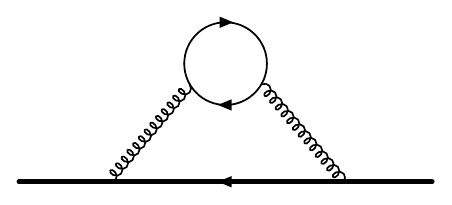}
\includegraphics[width=0.24\textwidth]{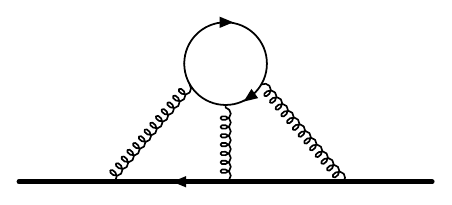}
\includegraphics[width=0.24\textwidth]{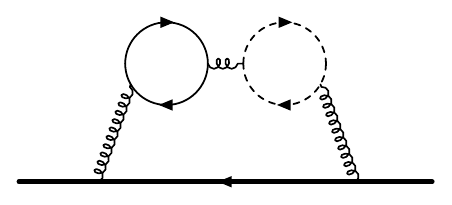}
\includegraphics[width=0.24\textwidth]{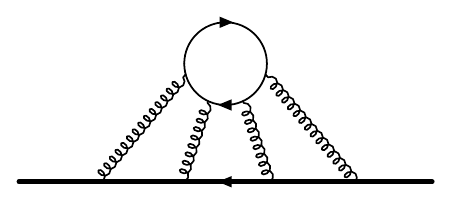}
\includegraphics[width=0.24\textwidth]{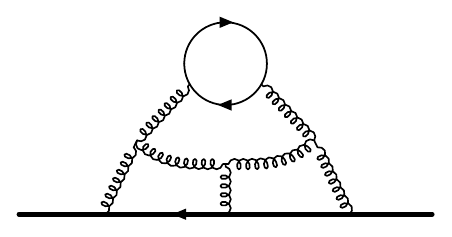}
\includegraphics[width=0.24\textwidth]{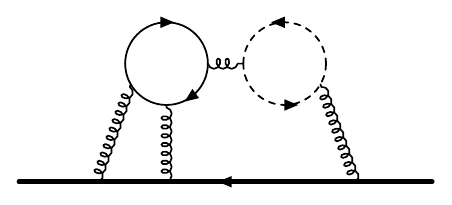}
\includegraphics[width=0.24\textwidth]{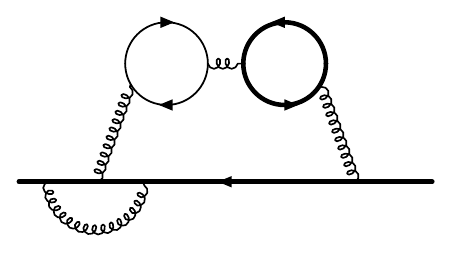}
\includegraphics[width=0.24\textwidth]{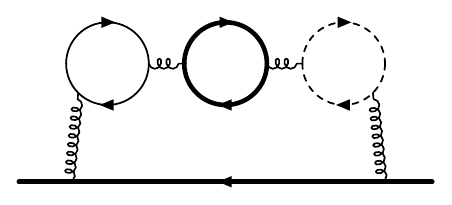}
\caption{\label{fig::sample_FDs}
 Examples of Feynman diagrams with two, three, and four loops
  required for the calculation of $Z_m^{\rm OS}$
  involving a second massive quark. Thick and thin solid lines
  denote quarks with masses $m_1$ and $m_2$; dashed lines represent massless quarks.
  Curly lines represent gluons.
  }
\end{figure}

For the computation of $Z_m^{\rm OS}$ to order $\alphas^4$
we have to consider the on-shell two-point function for heavy quarks
up to four loops. Representative Feynman diagrams can be found
in Fig.~\ref{fig::sample_FDs}.

\subsection{\label{sub::ampl}Amplitude}

To calculate the amplitude, we proceed as follows: We generate the
amplitudes for the heavy-quark two-point function using
\texttt{qgraf}~\cite{Nogueira:1991ex} and process them with
\texttt{tapir}~\cite{Gerlach:2022qnc} and \texttt{exp}~\cite{Harlander:1998cmq,Seidensticker:1999bb,exp_git} to generate
\texttt{FORM} code~\cite{Davies:2026cci}. Within \texttt{FORM}, we apply
projectors to the scalar and vector parts of the fermion self-energy,
evaluate the traces, and express each scalar integral in terms of a
function whose arguments represent the powers of the propagators. Each
of these functions belongs to one of 56~integral families.

In the next step, we reduce the list of integrals contributing to the amplitude to a basis of master integrals using integration-by-parts relations~\cite{Tkachov:1981wb,Chetyrkin:1981qh} and the Laporta algorithm~\cite{Laporta:2000dsw}.

In order to facilitate an efficient reduction, we first search for a basis in which the denominators of the reduction coefficients factorise in $\epsilon$ and $x$~\cite{Smirnov:2020quc,Usovitsch:2020jrk}.
In addition, we try to minimise the number of spurious poles in $\epsilon$.
To identify suitable candidates, we perform test reductions for each integral family separately, employing \texttt{Kira~2}~\cite{Maierhofer:2017gsa,Klappert:2020nbg} with \texttt{Fermat}~\cite{fermat} as reduction backend.
Afterwards, we use the script \texttt{ImproveMasters.m}~\cite{Smirnov:2020quc}\footnote{It was shipped together with \texttt{FIRE} from version \texttt{6.4.1} until version \texttt{6.6}~\cite{Smirnov:2023yhb}.} to select a suitable basis for each integral family.
With these bases at hand, we perform the full reduction for the integrals contributing to the amplitude (for Feynman gauge), again separating the integral families.
These reductions are carried out with \texttt{Kira~2}, this time employing \texttt{FireFly}~\cite{Klappert:2019emp,Klappert:2020aqs} for finite-field techniques~\cite{Kauers:2008zz,vonManteuffel:2014ixa,Peraro:2016wsq}.
In parallel, we perform an additional reduction involving only the candidate master integrals in order to identify relations and symmetries between different integral families and further minimise the basis.
The resulting minimal basis contains $562$ master integrals.


\subsection{\label{sub::eam}Expand-and-match}

To calculate the master integrals, we employ the ``expand-and-match''
method developed in Refs.~\cite{Fael:2021kyg,Fael:2021xdp,Fael:2022miw}. In fact, this
procedure was first applied to the calculation of eight of the sixteen
colour structures, which involve $\nm$~\cite{Fael:2021kyg}. Since an analytical calculation was also
possible for these cases, we were able to verify that
``expand-and-match'' provides approximately ten significant
digits across the entire kinematic range.

As a first step, we derive differential equations in $x$~\cite{Kotikov:1990kg,Remiddi:1997ny,Gehrmann:1999as}.
We employ \texttt{LiteRed}~\cite{Lee:2012cn,Lee:2013mka} to differentiate the master integrals with respect to $x$ and use reduction tables obtained by \texttt{Kira~2} to establish the differential equations.

The basic idea behind ``expand-and-match'' is to construct high-order
expansions -- comprising roughly 50 to 100 terms -- for each master integral
around suitably chosen expansion points. To this end, we substitute an
ansatz for the master integrals into the differential equation. This
leads to linear equations in which the coefficients of the ansatz are
expressed in terms of just a few boundary constants. These constants
are determined by matching the ansatz to known results for the master
integrals at a value of $x$ near the expansion point (or at the
expansion point itself). In some applications, the reference integrals
can be obtained via specialized numerical computations of the master
integrals, for instance using \texttt{AMFlow}~\cite{Liu:2017jxz,Liu:2021wks,Liu:2022mfb,Liu:2022chg}. In the present case,
however, this approach is too computationally intensive (or even
impossible with the current versions of the tools).
We therefore employ ``expand-and-match'' in the manner
introduced in Ref.~\cite{Fael:2021kyg}: We compute all master integrals
analytically in the limit $x \to \infty$ (see Section~\ref{sub::largex}) and then proceed stepwise to
smaller values of $x$. At each step, the boundary constants for the
new expansion are fixed by evaluating the previous expansion at a
value of $x$ where both expansions converge.

To cover the whole region for $x$ we find it convenient
to use the variable $s=x^2$ and choose the following expansion points:
\begin{eqnarray}
s_0 \in \left\{0,\frac{1}{32},\frac{1}{16}
\frac{1}{10},
\frac{1}{6},
\frac{1}{4},
\frac{1}{2},
\frac{3}{4},
1,
\frac{5}{4},
\frac{3}{2},2,3,4,6,8,12,18,20,35,\infty\right\}.
\end{eqnarray}
With the exception of the expansions around $s_0=0$ and $s_0=\infty$,
all of them are simple Taylor expansions,
and we can use the following ansatz for each master integral:
\begin{eqnarray}
  M_n &=& \sum\limits_{i=-4}^{i_{\rm max}} \sum\limits_{j=0}^{j_{\rm max}} c_{n,i,j} \epsilon^i (s-s_0)^j
  \,.
\end{eqnarray}
We can choose $-4$ as the lower limit of the $\epsilon$-expansion,
since poles at most of order $1/\epsilon^4$ can appear in a single
master integral.  The upper limit for $i$ is determined such that the
combination of the prefactor (stemming from the reduction) and the master integral includes all
terms up to order $\epsilon^0$.  The value $0$ can be chosen as the
lower limit for $j$; however, the upper limit depends on the order of
the spurious poles in $s-s_0$ within the coefficient of the respective
master integral.

For $s_0=0$ and $s_0=\infty$ the ansatz for the individual master
integrals has to contain also logarithms and we choose for $s\to0$
\begin{eqnarray}
  M_n &=& \sum\limits_{i=-4}^{i_{\rm max}}
  \sum\limits_{j=j_{\rm min}}^{j_{\rm max}}
  \sum\limits_{k=0}^{i+4}
  c_{n,i,j,k} \epsilon^i {s^{j/2} (\log(s))^k}
  \,,
\end{eqnarray}
and for $s\to \infty$
\begin{eqnarray}
  M_n &=& \sum\limits_{i=-4}^{i_{\rm max}}
  \sum\limits_{j=j_{\rm min}}^{j_{\rm max}}
  \sum\limits_{k=0}^{i+4}
  c_{n,i,j,k} \epsilon^i {\frac{1}{s^j} (\log(s))^k}
  \,.
\end{eqnarray}
In both cases the lower limit of $k$ is zero and the upper limit
is $i+4$.

The expansion around $s=1$ is rather peculiar.
The master integrals have to be regular at the point $s=1$, however, the differential equations develop high poles in $1/(1-s)$ (of up to degree 9).
In order to reduce the number of unknown boundary conditions at $s=1$, we do a dedicated reduction of the master integrals at this point and find a reduced set of 210 master integrals.
This reduction can be used to relate boundary conditions in the expand-and-match approach or to check the results obtained.

As an additional check, we use \texttt{AMFlow} to compute numerical values of the master integrals at $s=1$.
We note, that we were not able to compute numerical values for all master integrals this way.

\subsection{\label{sub::largex}Four-loop result in the large-$x$ limit}

In the limit of large $m_2$, $Z_m^{\rm OS}$ can be calculated using
vacuum integrals up to four-loop order and on-shell integrals up to
three-loop order -- quantities that have already been extensively studied
in the literature (see, e.g.,
Refs.~\cite{Lee:2010ik,Schroder:2005va,Chetyrkin:2015mxa}). Indeed, all
the necessary components are available in analytic form. It is
therefore natural to choose the limit $s\to\infty$ as the starting
point for the ``expand-and-match'' procedure. In the following, we
briefly describe our calculation.

Our goal is to calculate all master integrals for the case $m_2 \gg
m_1$.
An essential tool for implementing this limit is
\texttt{exp}~\cite{Harlander:1998cmq,Seidensticker:1999bb,exp_git}, which automates the
hard-mass procedure (see, e.g., Ref.~\cite{book_smirnov_2012}).
It operates at the level of individual diagrams and
generates \texttt{FORM} code~\cite{Davies:2026cci} that is easily processed
further.

The asymptotic expansion of four-loop on-shell master integrals with a
through-going massive line leads either to four-loop vacuum integrals,
whose sole mass scale is given by $m_2$, or to a factorization of
$n$-loop vacuum integrals and $(4-n)$-loop on-shell integrals with
$n=1, 2, 3$. In the (private) code \texttt{calc}, the calculation of
such integral products is automated, including the necessary tensor
reduction prior to the initial integration, which, in our case,
involves vacuum integrals. The output of \texttt{calc} consists of
results for each master integral, expressed as products of two scalar
functions belonging to a vacuum or on-shell integral family. In the
case of four-loop vacuum integrals, there is no second scalar
integral. We use \texttt{FIRE}~\cite{Smirnov:2023yhb} for the reduction to
master integrals.

Note that due to spurious poles in $\epsilon$ we need some of the
on-shell and vacuum master integrals to higher order in $\epsilon$.
Fortunately, up to the loop order we need, numerical results can be obtained with the program \texttt{SummerTime}~\cite{Lee:2015eva}.
Needed coefficients up to weight 9 are reconstructed with the \texttt{PSLQ}~\cite{pslq} algorithm.
Coefficients of even higher weight are left symbolic in the large mass expansion, they drop out of the physical results.
For the expand-and-match procedure, which requires a fully numerical input, we use results with up to 10000 digits.

We have calculated the first  two terms in $(m_1/m_2)^2$
for each master integral.
To fix the boundary conditions, we require approximately 40\% of the
information contained in the leading terms of all master integrals. The results for the
remaining expansion terms, specifically the subleading terms in
$(m_1/m_2)^2$, are reproduced from the solution to the differential
equation, providing a rigorous consistency check of our calculation.

\section{\label{sec::ren}Renormalization}

In this Section we describe the renormalization of $Z_m^{\rm OS}\equiv
Z_{m_1}^{\rm OS}$ to four loops including the $\nm$ contributions.
For the one- to three-loop expressions we use the results from
Refs.~\cite{Marquard:2015qpa,Marquard:2016dcn,Fael:2020bgs} which
contain all relevant higher order terms in $\epsilon$. This is also
the case for the three-loop $\nm$ contribution computed in
Ref.~\cite{Fael:2020bgs}.  We only have to renormalize the strong
coupling constant $\alphas$, which we do in the $\overline{\rm MS}$
scheme, and the mass parameters $m_1$ and $m_2$, which we renormalize
on-shell.

The counterterm for $\alphas$ is needed up to three loops and can,
e.g., be found in Ref.~\cite{Chetyrkin:2004mf}.  The $\overline{\rm
MS}$ renormalization constant, $Z_{\alphas}$, depends on the number of
flavours, $\nf$, for which we use $\nf = \nl + \nm + \nh$. After
multiplying the one-, two-, and three-loop results with $Z_{\alphas}$ we
can drop all four-loop terms which do not involve $\nm$.

To renormalize the masses $m_1$ and $m_2$ we generate together with
the one-, two-, and three-loop amplitudes the corresponding mass
counterterm insertions. Then, the bare amplitude contains the
respective renormalization constants, $Z_{m_1}^{\rm OS}$ and
$Z_{m_2}^{\rm OS}$, which we simply have to replace taking into
account the proper loop-order.

In the bare one-loop amplitude only $Z_{m_1}^{\rm OS}$ appears, which
is needed to three-loop order, including the $\nm$ terms up to order
$\epsilon$~\cite{Fael:2020bgs}.  $Z_{m_2}^{\rm OS}$ appears for the
first time at two loops and thus at most two-loop corrections are
needed. Note that all colour structures have to be included.  The
two-loop amplitude also contains $Z_{m_1}^{\rm OS}$, which is required
here as well -- incorporating all colour factors -- up to two loops.
Since at one-loop order the mass renormalization constant does not
have a $\nf$ dependence, it is sufficient to consider the $\nm$ and
$\nm^2$ contributions in the bare three-loop amplitude.  It contains
both $Z_{m_1}^{\rm OS}$ and $Z_{m_2}^{\rm OS}$ for which we have to
use the one-loop expressions.

\begin{figure}[t]
\centering
\begin{tabular}{ccc}
\includegraphics[width=0.3\textwidth]{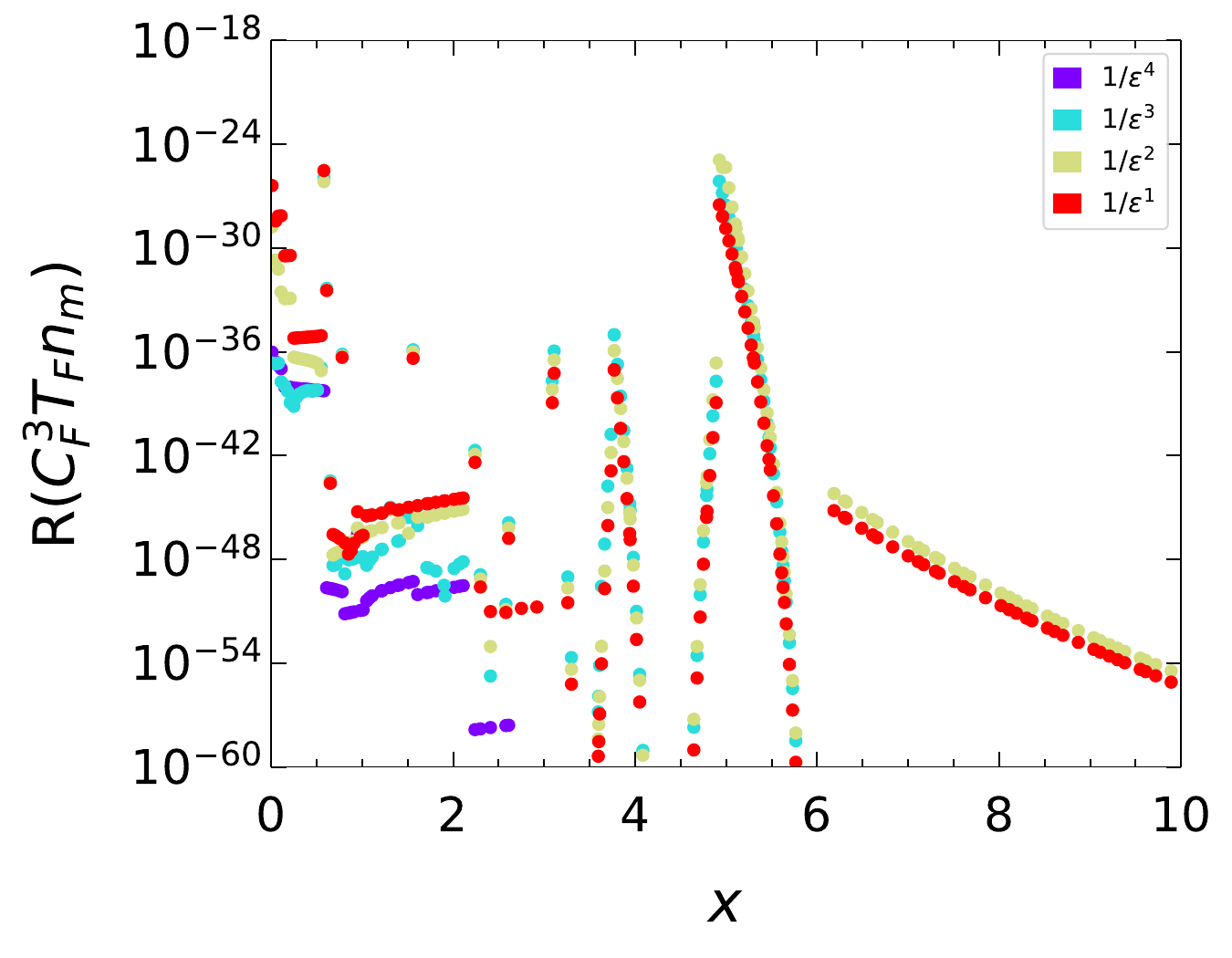}&
\includegraphics[width=0.3\textwidth]{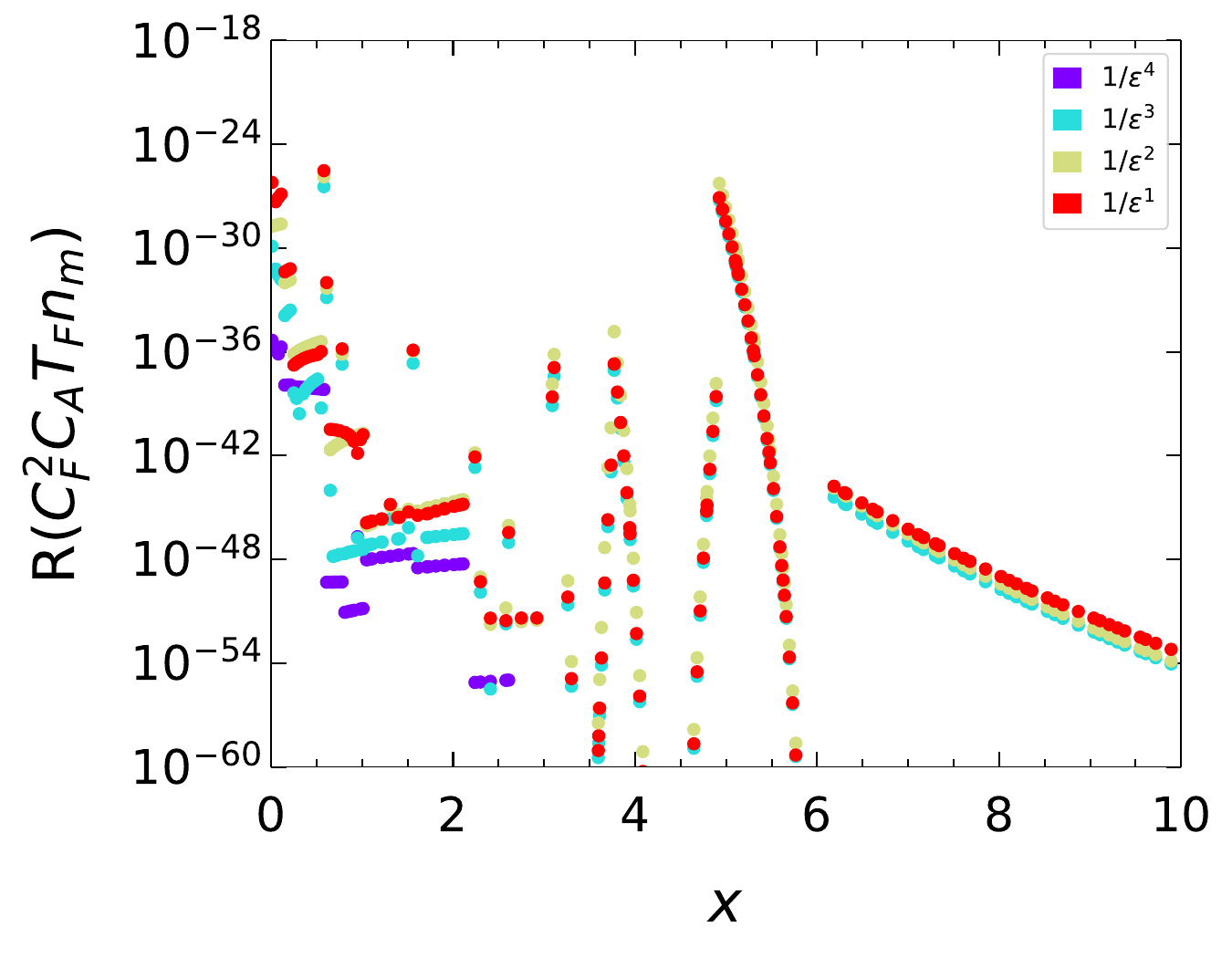}&
\includegraphics[width=0.3\textwidth]{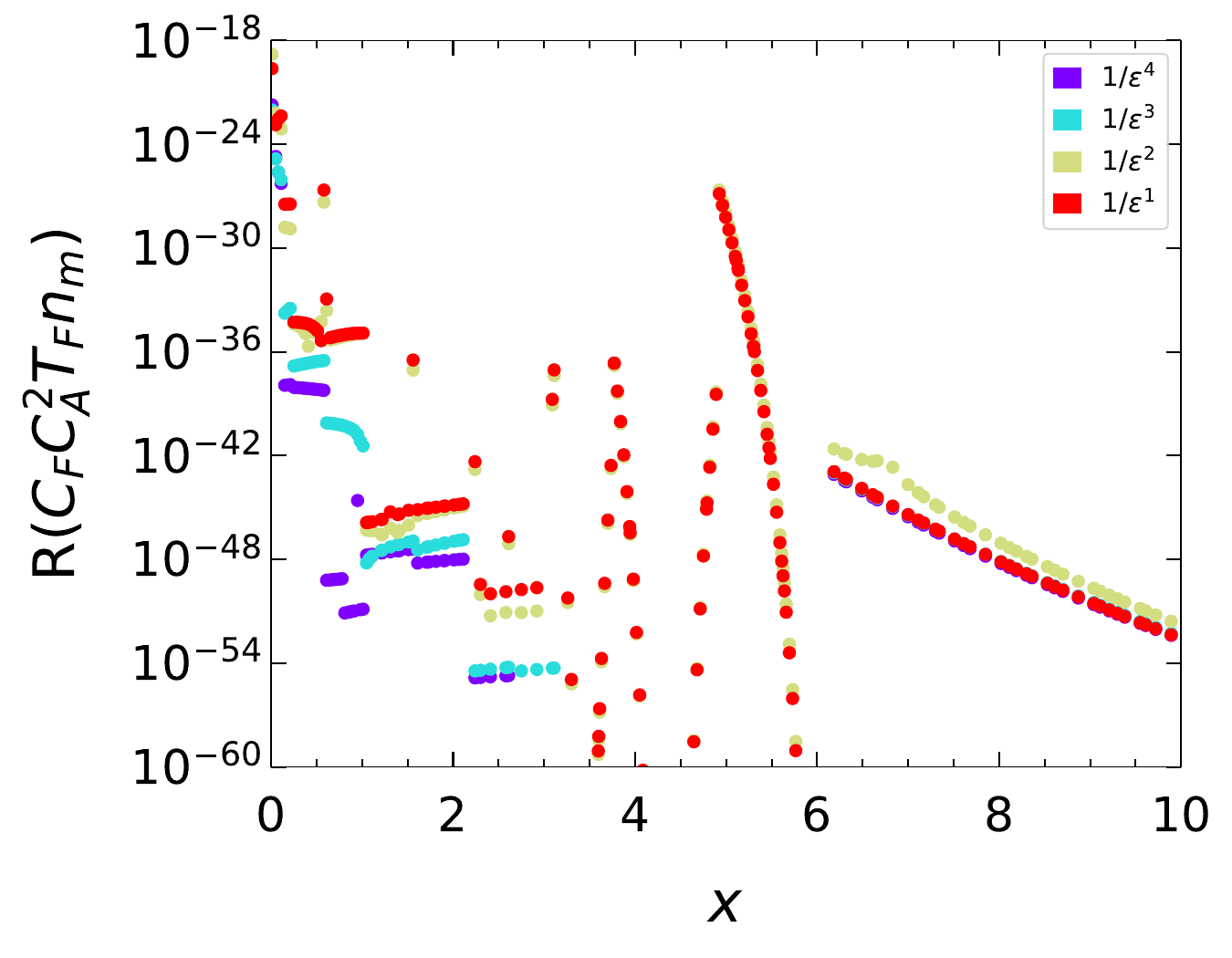}
\\
\includegraphics[width=0.3\textwidth]{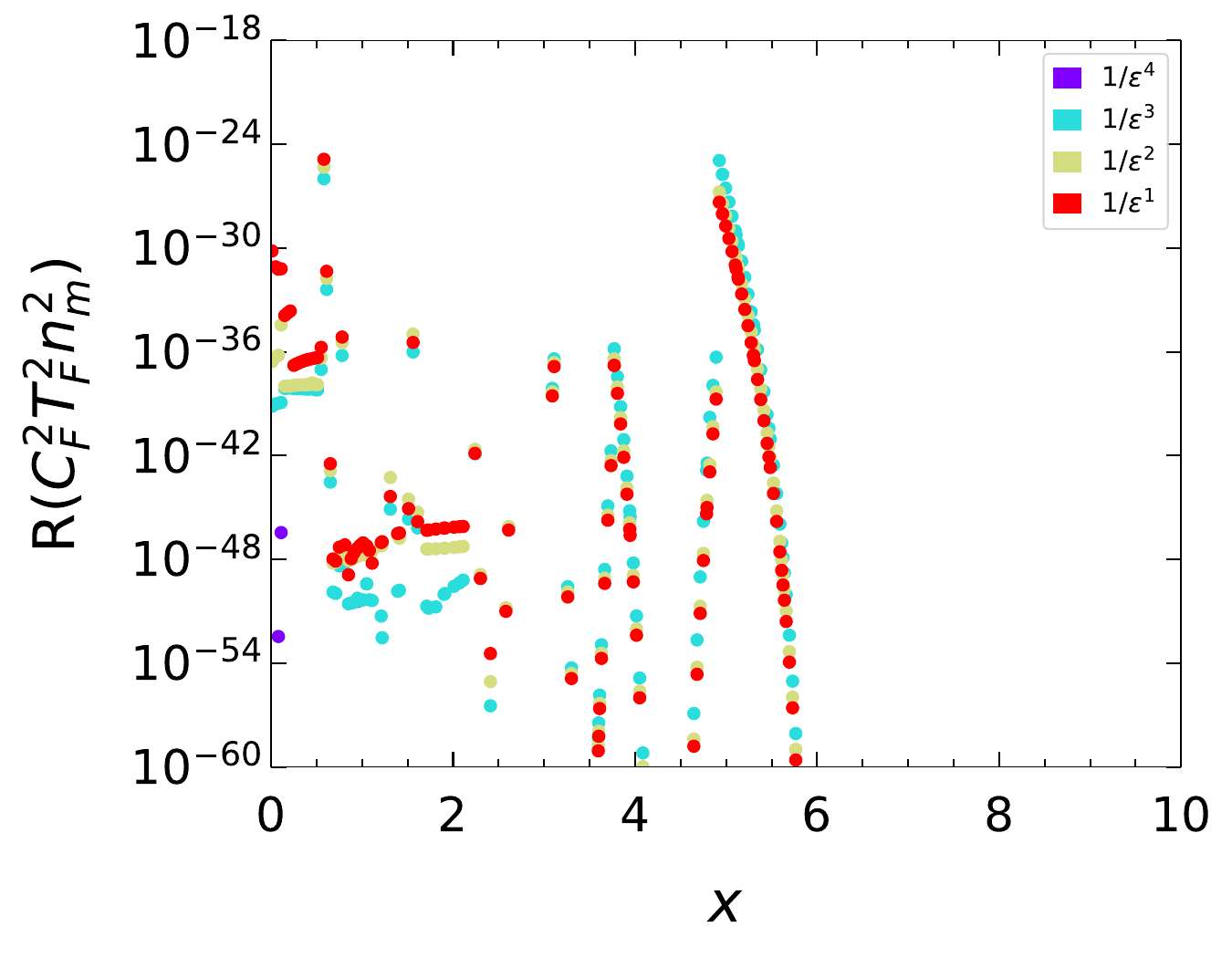} &
\includegraphics[width=0.3\textwidth]{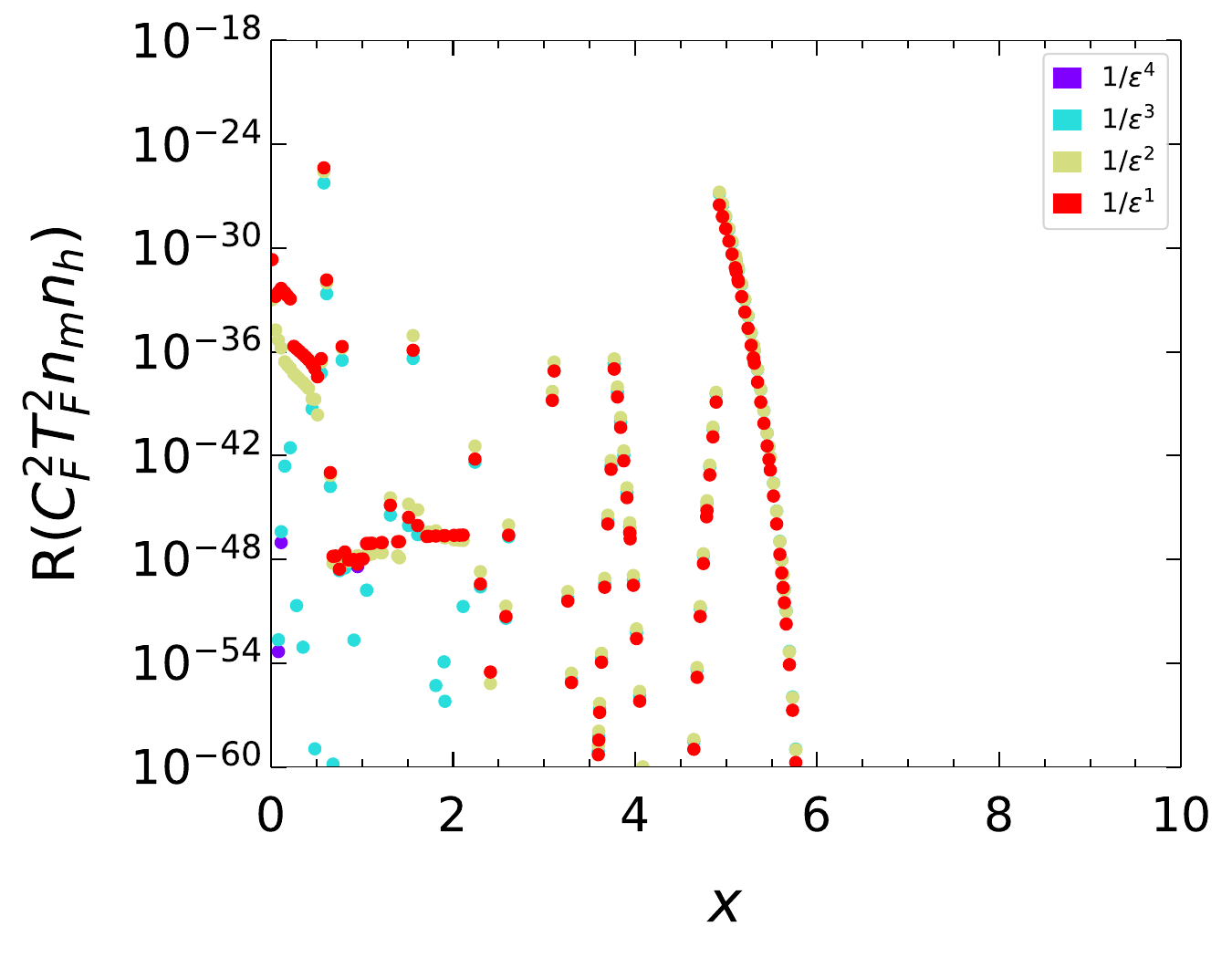}&
\includegraphics[width=0.3\textwidth]{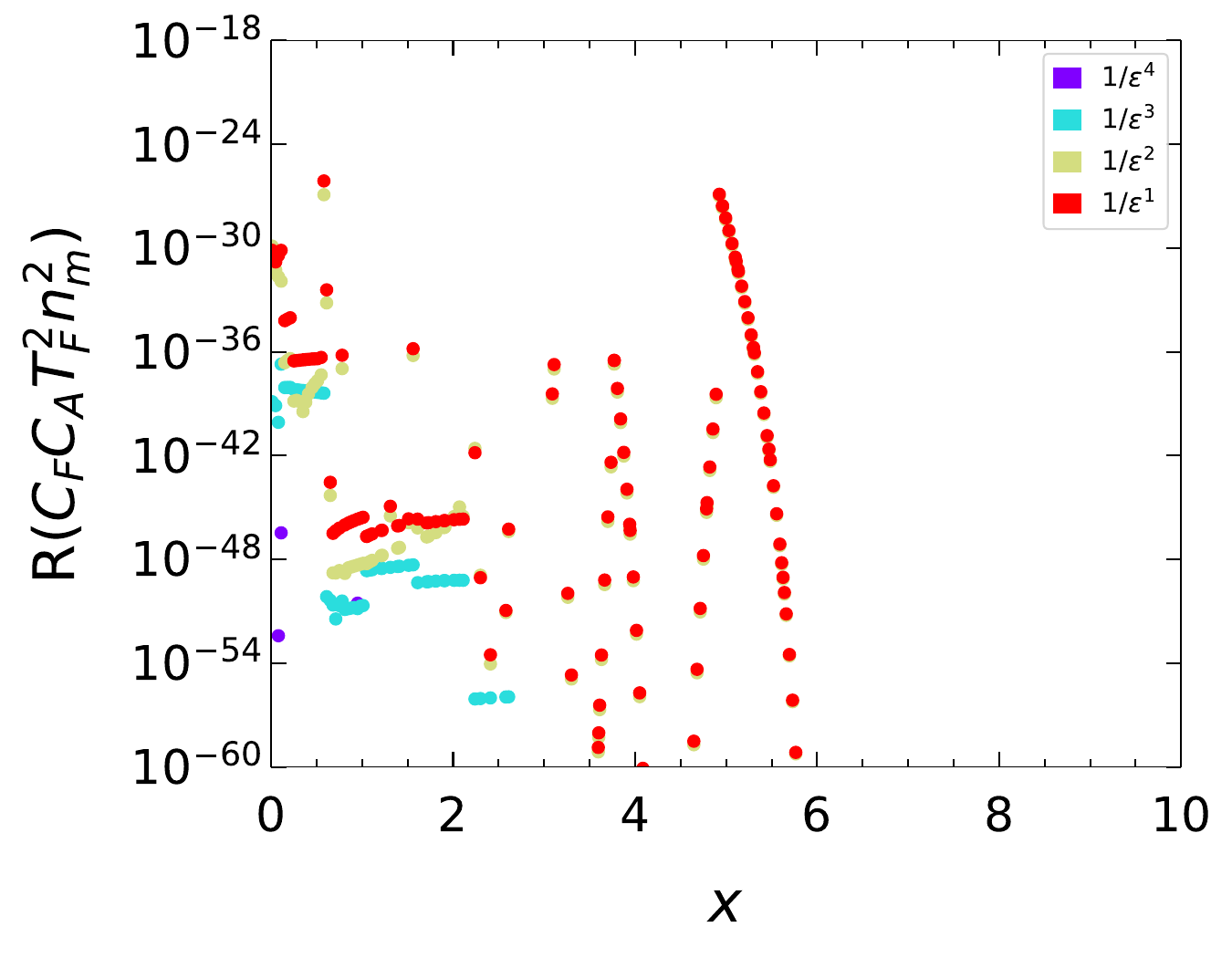}
\\
\includegraphics[width=0.3\textwidth]{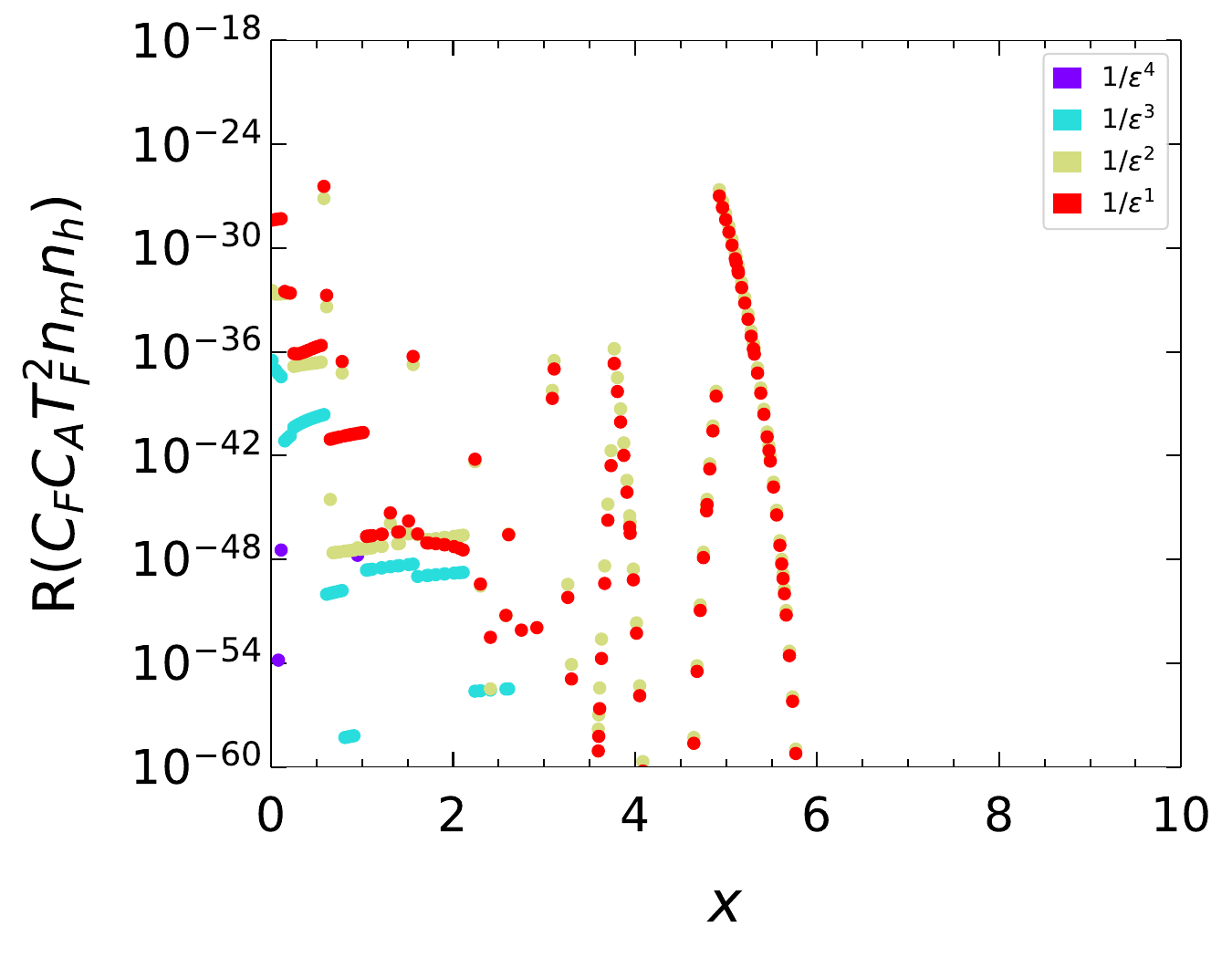}
\end{tabular}
\caption{\label{fig::pole_cancel}
  Pole cancellation plots  for the
  seven colour structures which are not known analytically. The eighth colour structure is finite.
  In each case we show the quantity $R$
  from Eq.~(\ref{eq::R}) for the poles
  $1/\epsilon^i$ with $i=1,2,3,4$.
  }
\end{figure}

All our results up to three-loop order are available analytically,
including the two-mass contribution at order $\epsilon$.    On the other hand, the four-loop result is
known -- as described in the previous section -- as generalized
expansions with numerical coefficients.  Consequently, the poles in
$\epsilon$ within the finite quantity $z_m$ cancel out only
numerically.  We can utilize this cancellation to estimate the
numerical accuracy of the method we apply to the four-loop amplitude.

$Z_m^{\rm OS}$ contains ``physical'' poles of at most fourth order.
At four-loop order we also observe spurious poles which are
introduced by the integration-by-parts procedure.  These poles cancel
out after substituting the explicit results for the master integrals. In our
calculation we have spurious poles up to order
$1/\epsilon^{10}$. We find that across the entire $x$-range
the coefficients of $1/\epsilon^5$, \ldots, $1/\epsilon^{10}$ are
at most about $10^{-40}$. Regarding the ``physical'' poles, we discuss their
cancellation in Fig.~\ref{fig::pole_cancel}; this figure shows, for each
colour structure and each $1/\epsilon^k$ pole with $k=1,2,3,4$, the magnitude
\begin{eqnarray}
        R_i(C)&=&
       \left| \frac{ z_m^{(4),\rm bare} \Big|_{C,\epsilon^i}
          + z_m^{(4),\rm CT} \Big|_{C,\epsilon^i}}
          { z_m^{(4),\rm CT} \Big|_{C,\epsilon^i} }
        \right|\,,
        \label{eq::R}
\end{eqnarray}
where $z_m^{(4),\rm bare}$ is the bare four-loop result and
$z_m^{(4),\rm CT}$ the counterterm contribution discussed at the
beginning of this section. $C$ denotes one of the colour structures in Eq.~(\ref{eq::zm4_col})
and the notation ``$C,\epsilon^i$'' in the subscript indicates that the coefficient of
the colour factor $C$ and $1/\epsilon^i$
is being considered. The results for the seven colour structures that are not analytically
known and that are divergent are shown in the sub-figures.\footnote{Note
that $\zmdFFM$ is finite.} We initiate the ``expand-and-match''
procedure at $x \to \infty$, where we perform the matching against the
exact result. Subsequently, we proceed to smaller values of $x$,
following the algorithm outlined in Section~\ref{sub::eam}. This implies
that the accuracy of the individual master integrals cannot improve as
we move towards $x=0$. However, in Fig.~\ref{fig::pole_cancel}, we
observe significant fluctuation in the accuracy with which the
$\epsilon$-poles cancel out, particularly in the range $2 \lesssim x
\lesssim 6$. This behaviour can be explained by the very high accuracy
of the master integrals combined with the re-expansion of the
amplitude at each intermediate expansion point. This leads to a
rearrangement of the various $\epsilon$-terms and results in an
$x$-dependent fluctuation in the accuracy of the amplitude.

It should be noted that not all curves are
displayed in the individual panels, as we truncate at $10^{-60}$ and, in
many cases, pole cancellation occurs to an even higher number of digits.

The contributions involving more closed fermion loops originate from
Feynman diagrams that exhibit a topologically simpler structure. This
is reflected in a stronger cancellation of the $1/\epsilon$
poles. Indeed, for contributions with two or three
(not shown explicitly)
closed fermion
loops, approximately 25 digits cancel each other out in the worst-case
scenario. For the three colour
structures involving a single closed quark loop, the cancellation is
(somewhat) less pronounced in the region around $x \to 0$,
though we still observe a cancellation of at least 18 digits. For
values of $x \gtrsim 6 $, the cancellation is significantly stronger,
exceeding 40 digits for all colour factors. In this region
there is hardly any difference between the various pole
terms.
For $x<1$, we in general observe improved cancellation for the higher pole
terms.
In the region $3.5\lesssim x\lesssim 6$ the $1/\epsilon$ and
$1/\epsilon^2$ terms cancel with approximately the
same quality whereas for the higher poles
the cancellation is significantly better (and not shown
in the plots).

Based on the quality of the pole cancellation, as shown in
Fig.~\ref{fig::pole_cancel}, we estimate that at least 15 significant
digits are preserved in the finite term of $z_m$.
It should be noted that, for the finite $\epsilon$-term, a cancellation of about one significant digit occurs between $z_m^{(4),\rm bare}$ and $z_m^{(4),\rm CT}$.

\section{\label{sec::results}Results and applications}

In this section, we present the results of our computation.
We split the presentation into the following parts.
In subsection \ref{sub::res_gen}, we present results for $\zmM$ as well as leading expansion terms around the physically interesting expansion points $x \to 0$, $x \to 1$ and $x \to \infty$.
In \ref{sub::singlemass}, we compare the numerical values in the limits $x = 0$ and $x=1$ to the values previously obtained in Refs.~\cite{Marquard:2015qpa,Marquard:2016dcn}.
Subsequently, we discuss the size of the power corrections in $x$ at the various loop orders and compare our results with a prediction of the linear power correction made in Refs.~\cite{Hoang:2017btd,Mateu:2017hlz}.
In subsection \ref{sub::rundec}, we present the numerical implementation into the public packages \texttt{RunDec} and \texttt{CRunDec}.

\subsection{\label{sub::res_gen}General case}

\begin{figure}[t]
    \centering
    \begin{tabular}{c}
      \includegraphics[width=0.95\linewidth]{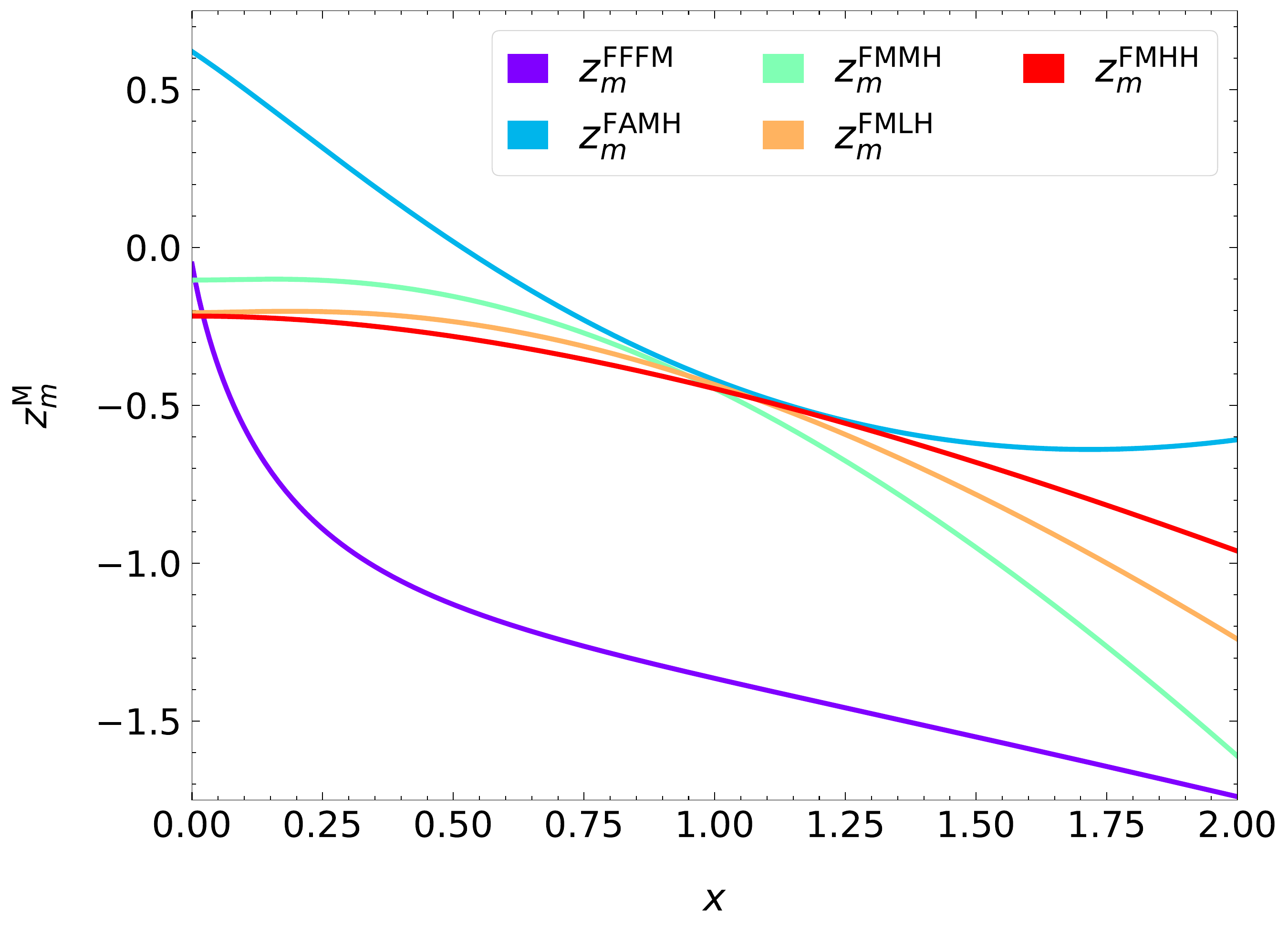}
    \end{tabular}
    \caption{Five of the 16 four-loop results for $\zmM$ as a function of $x=m_2/m_1$.
    }
    \label{fig::zM1}
\end{figure}

\begin{figure}[t]
    \centering
    \begin{tabular}{c}
        \includegraphics[width=0.95\linewidth]{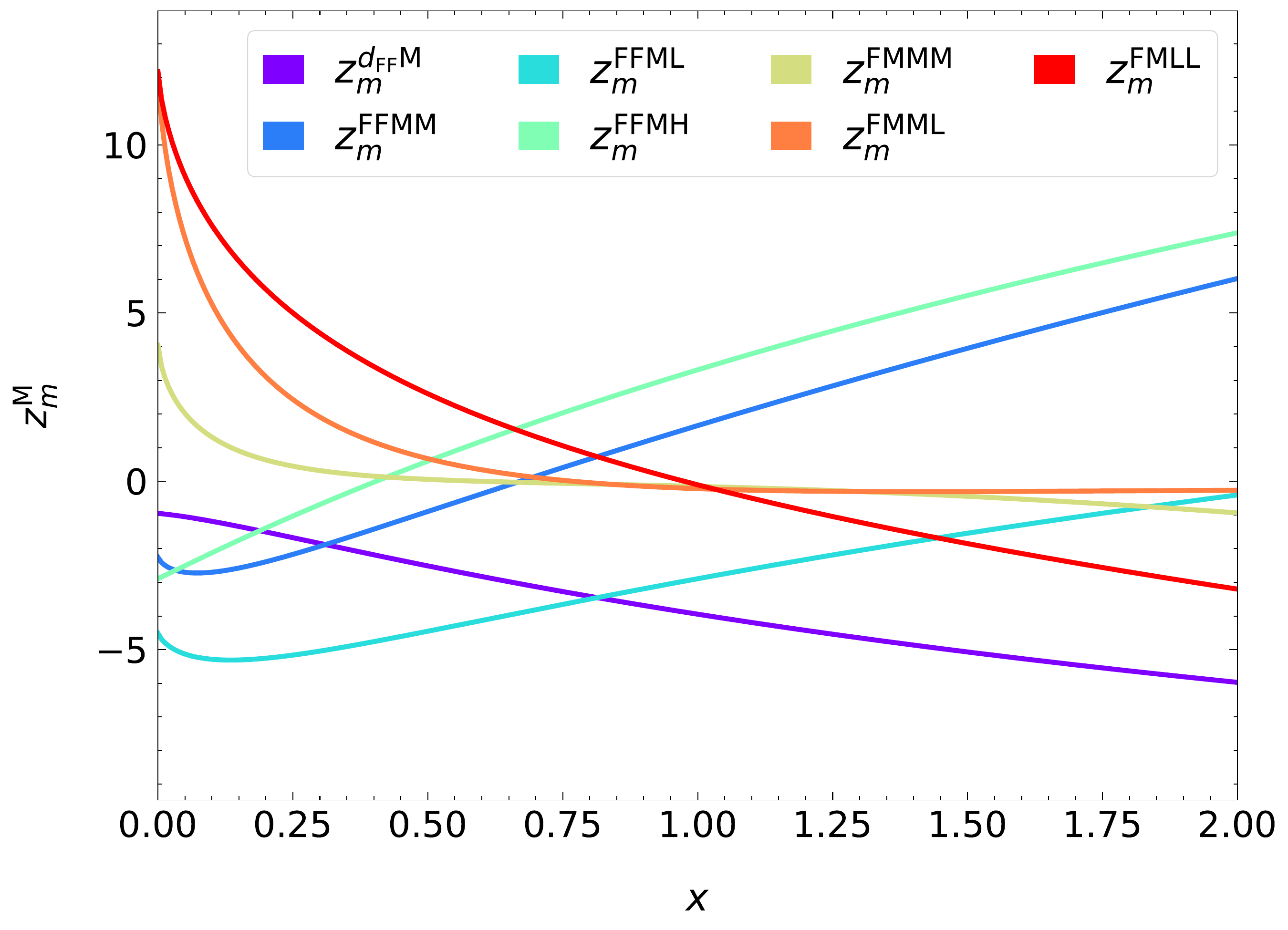}
    \end{tabular}
    \caption{Continuation of Fig.~\ref{fig::zM1}: Seven
     four-loop colour structures for $\zmM$ as a function of $x=m_2/m_1$. }
    \label{fig::zM2}
\end{figure}

\begin{figure}[t]
    \centering
    \begin{tabular}{c}
    \includegraphics[width=0.95\linewidth]{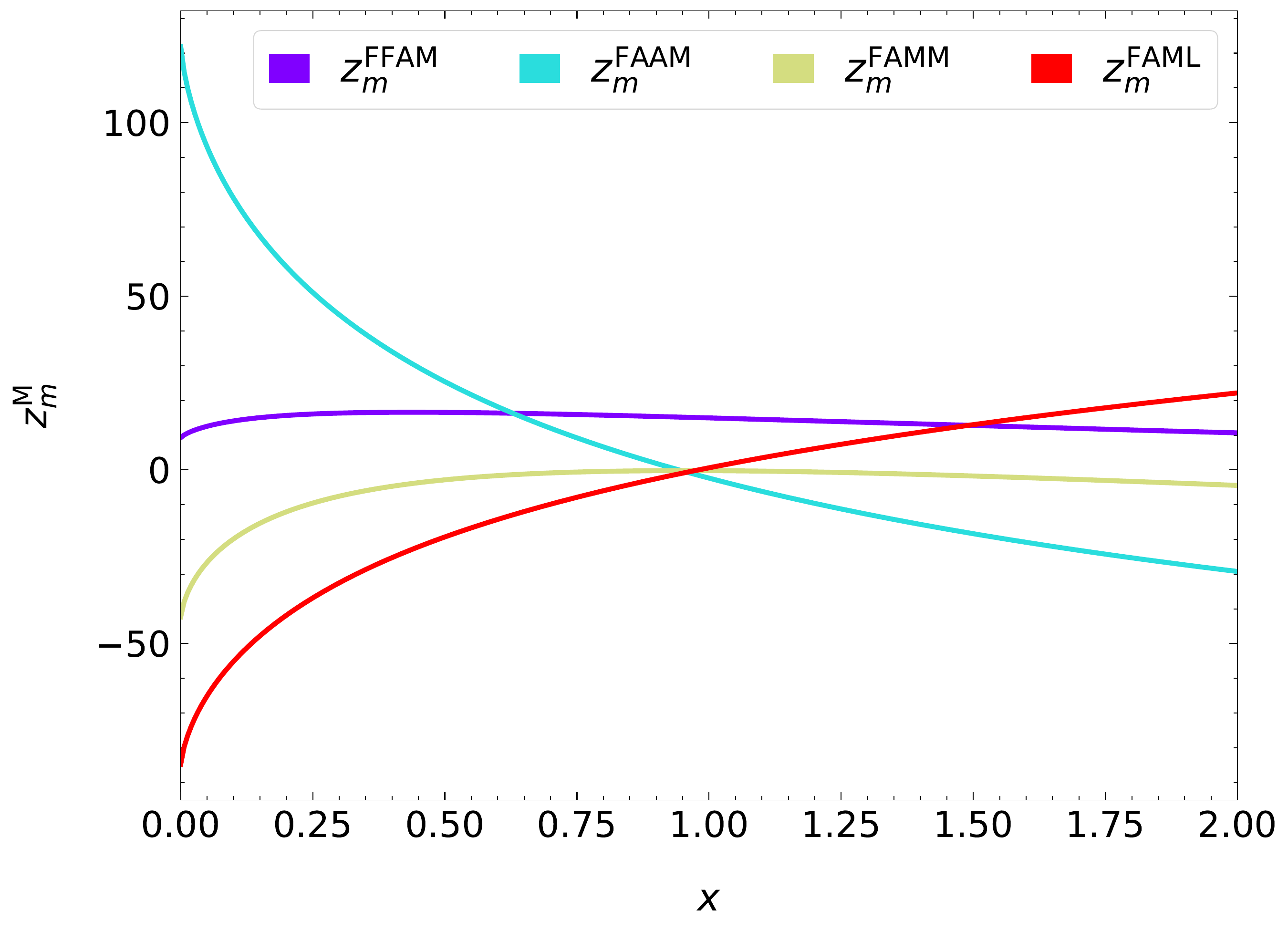}
    \end{tabular}
    \caption{Continuation of Fig.~\ref{fig::zM2}: The four four-loop results for $\zmM$ which show the strongest variation
    as a function of $x=m_2/m_1$.}
    \label{fig::zM3}
\end{figure}

Our main results are presented in Figs.~\ref{fig::zM1},~\ref{fig::zM2} and~\ref{fig::zM3}.
We show
the coefficients of the 16 individual colour structures of $\zmM$ as a
function of $x$ for $0 \le x \le 2$. We have divided the coefficients
into three groups: the five
functions whose magnitude varies by at most about 1 for $x \in [0,
2]$; the seven functions for which $|\zmM| \lesssim 10$ holds; and
the remaining four functions, which exhibit stronger variation. For
some of the functions, we observe strong variation as $x \to
0$. However, it should be noted that there is no logarithmic
divergence, and the limit as $x \to 0$ exists in all cases. The
situation is different for $x \to \infty$ (not shown in
Figs.~\ref{fig::zM1},~\ref{fig::zM2} and~\ref{fig::zM3}); in this limit, we observe $\log^n(x)$ terms with
$n \le 4$.

There are several pairs of functions whose quotient is an integer for
$x = 0$ and $x = 1$, respectively.
These relations simply follow from setting quark masses equal.
For $x=0$, the quarks with mass $m_2$ become equivalent to the massless quarks and the diagrams agree identically.
However, one has to take into account the number of possible diagram permutations before setting the masses equal.
For example, both $\zmFMLL$ and $\zmFMML$ have three permutations of every diagram each, and, thus, one finds $\zmFMLL(x=0) = \zmFMML(x=0)$.
On the other hand, each diagram comes in two permutations for $\zmFAML$, whereas there is only one permutation for $\zmFAMM$, and one has $\zmFAML(x=0) = 2 \zmFAMM(x=0)$.
Similarly, for $x = 1$ the quarks with mass $m_2$ become equivalent to the quarks with mass $m_1$ and relations like $\zmFFMH(x=1) = 2 \zmFFMM(x=1)$ follow.
These relations can be observed in the plots shown in
Figs.~\ref{fig::zM1},~\ref{fig::zM2}, and~\ref{fig::zM3}.
We have verified that all these relations hold with high
precision -- specifically, to about 30 significant digits.

Looking at Fig.~\ref{fig::zM3}, it appears as though the three
functions $\zmFAAM$, $\zmFAMM$, and $\zmFAML$ intersect at an
$x$-value close to $x=1$. However, closer inspection reveals that
pairs of curves intersect at $x$-values that are close to one another.
In any case, there is no obvious relationship between these three
functions.

\begin{figure}[t]
    \centering
    \includegraphics[width=0.95\linewidth]{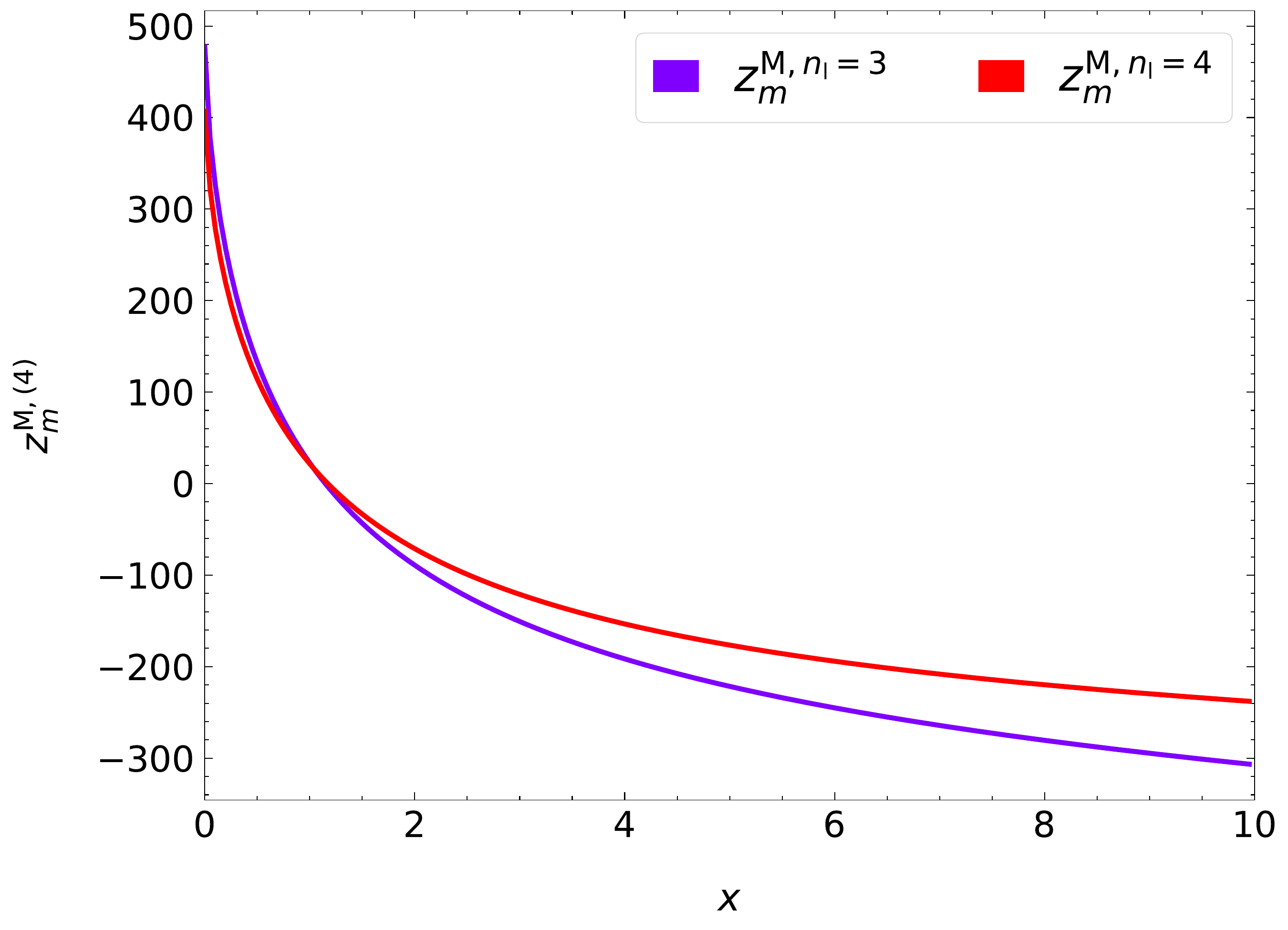}
    \caption{The complete four-loop contributions for $\zmM$ for $\nl=3$ and $\nl=4$ massless
    quark flavours.}
    \label{fig::zM_tot}
\end{figure}

In Fig.~\ref{fig::zM_tot}, we show $\zmM$ as a function of $x$ for $x
\in [0, 10]$ after substituting the numerical values for the colour
factors. We present the results for $\nl=3$ and $\nl=4$. We observe a
strong variation in $\zmMfour$, ranging from $\zmMfour =
\mathcal{O}(400)$ at $x=0$ to a value below $-200$ at $x=10$. At
$x=1$, $\zmMfour$ is numerically small
and only of order~10.

For $\nl=3$ and $x\approx0.3$, we have $m_1=m_b$. In this case, the
four-loop corrections to $\zmM$ are smaller by a factor of about two
than in the $m_c=0$ approximation. The approximation $m_c=m_b$ (i.e.,
$x=1$) would also be a poor approximation. For $\nl=4$ and
$x\approx 0.02$,  corresponding to $m_1=m_t$, we observe a reduction in
$\zmMfour$ of about 10\% compared to the $m_b=0$
approximation. This demonstrates that the contribution of light-quark
mass effects becomes increasingly significant at higher loop orders.

For all expansion points, we have calculated at least 50 expansion
terms. For illustration we present below the first three expansion
terms for small and large values of $m_2$
and for $m_1\approx m_2$,
for both $\zmM(x)$ and
$\ymM(x_{fq})$. To obtain compact expressions, we substitute
numerical values for all colour factors.

For $x\to0$ we have

\begin{eqnarray}
\zmM/\nm &=& \left( \frac{\alphas}{\pi} \right)^2
\biggl\{
    +1.041367
    -1.644934 x
    + x^2
\biggr\}
 + \left( \frac{\alphas}{\pi} \right)^3
\biggl\{
            +26.94609
            \nonumber \\ &&
            -0.02224348\nh
            -1.305381\nl
            -0.6526907\nm
    + x
    \biggl(
            -21.38458
            \nonumber \\ &&
            +1.038544\nl
            +0.2924327\nm
        + l_x
        \biggl[
            +18.09427
            -1.096623\nl
            \nonumber \\ &&
            -1.096623\nm
        \biggr]
    \biggr)
    + x^2
    \biggl(
            +13.14653
            -0.2222222\nh
            -0.2222222\nl
            \nonumber \\ &&
            -0.2222222\nm
        + l_x
        \biggl[
            +10.68698
        \biggr]
    \biggr)
\biggr\}
 + \left( \frac{\alphas}{\pi} \right)^4
\biggl\{
            +757.6456
            -0.6646237\nh
            \nonumber \\ &&
            -0.03617151\nh^2
            -86.93043\nl
            -0.0344042\nh\nl
            +2.034423\nl^2
            -43.46522\nm
            \nonumber \\ &&
            -0.01720209\nh\nm
            +2.034423\nl\nm
            +0.678141\nm^2
    + x
    \biggl(
            -809.5624
            \nonumber \\ &&
            +2.317914\nh
            +98.62834\nl
            -2.432721\nl^2
            +56.95064\nm
            -2.801614\nl\nm
            \nonumber \\ &&
            -1.067526\nm^2
        + l_x
        \biggl[
            +388.7599
            -43.36398\nl
            +1.038544\nl^2
            -31.05313\nm
            \nonumber \\ &&
            +1.330977\nl\nm
            +0.2924327\nm^2
        \biggr]
        + l_x^2
        \biggl[
            -149.2778
            +18.09427\nl
            \nonumber \\ &&
            -0.5483114\nl^2
            +18.09427\nm
            -1.096623\nl\nm
            -0.5483114\nm^2
        \biggr]
    \biggr)
    \nonumber \\ &&
    + x^2
    \biggl(
            +447.7757
            -3.517769\nh
            -0.04695292\nh^2
            -49.5443\nl
            \nonumber \\ &&
            +0.1090961\nh\nl
            +1.002015\nl^2
            -54.27897\nm
            +0.1090961\nh\nm
            \nonumber \\ &&
            +2.00403\nl\nm
            +1.002015\nm^2
        + l_x
        \biggl[
            +91.55747
            -0.8888889\nh
            -4.127671\nl
            \nonumber \\ &&
            -4.127671\nm
        \biggr]
        + l_x^2
        \biggl[
            -44.79364
            +4.010158\nl
            +4.010158\nm
        \biggr]
    \biggr)
\biggr\}
\nonumber \\ &&
+ \mathcal{O}(x^3,\alphas^5)~,
\end{eqnarray}
where $x=m_2^{\rm OS}/m_1^{\rm OS}$, $l_x=\log(x)$,
and for the renormalization scale $\mu^2=(m_1^{\rm OS})^2$ has been chosen.
For the inverse relation we obtain in the small-$m_2$ limit
\begin{eqnarray}
\ymM/\nm &=& \left( \frac{\alphas}{\pi} \right)^2
\biggl\{
    -1.041367
    +1.644934 x_{fq}
    - x_{fq}^2
\biggr\}
 + \left( \frac{\alphas}{\pi} \right)^3
\biggl\{
            -26.67738
            \nonumber \\ &&
            +0.02224348\nh
            +1.305381\nl
            +0.6526907\nm
    + x_{fq}
    \biggl(
            +22.4812
            \nonumber \\ &&
            -1.038544\nl
            -0.2924327\nm
        + l_{x_{fq}}
        \biggl[
            -21.38414
            +1.096623\nl
            \nonumber \\ &&
            +1.096623\nm
        \biggr]
    \biggr)
    + x_{fq}^2
    \biggl(
            -13.8132
            +0.2222222\nh
            +0.2222222\nl
            \nonumber \\ &&
            +0.2222222\nm
        -6.686982 l_{x_{fq}}
    \biggr)
\biggr\}
 + \left( \frac{\alphas}{\pi} \right)^4
\biggl\{
            -744.8564
            -0.9031168\nh
            \nonumber \\ &&
            +0.03617151\nh^2
            +86.7581\nl
            +0.03440417\nh\nl
            -2.034423\nl^2
            +43.37905\nm
            \nonumber \\ &&
            +0.01720209\nh\nm
            -2.034423\nl\nm
            -0.678141\nm^2
    + x_{fq}
    \biggl(
            +822.7892
            \nonumber \\ &&
            +0.6729799\nh
            -100.0965\nl
            +2.432721\nl^2
            -55.86952\nm
            +2.801614\nl\nm
            \nonumber \\ &&
            +1.067526\nm^2
        + l_{x_{fq}}
        \biggl[
            -428.9246
            +1.188008\nh
            +45.16691\nl
            -1.038544\nl^2
            \nonumber \\ &&
            +30.17583\nm
            -1.330977\nl\nm
            -0.2924327\nm^2
        \biggr]
        + l_{x_{fq}}^2
        \biggl[
            +197.8033
            \nonumber \\ &&
            -0.5483114\nh
            -20.83583\nl
            +0.5483114\nl^2
            -20.28752\nm
            +1.096623\nl\nm
            \nonumber \\ &&
            +0.5483114\nm^2
        \biggr]
    \biggr)
    + x_{fq}^2
    \biggl(
            -455.9142
            +1.847672\nh
            +0.04695292\nh^2
            \nonumber \\ &&
            +50.16407\nl
            -0.1090961\nh\nl
            -1.002015\nl^2
            +52.40397\nm
            \nonumber \\ &&
            -0.1090961\nh\nm
            -2.00403\nl\nm
            -1.002015\nm^2
        + l_{x_{fq}}
        \biggl[
            +1.444618
            \nonumber \\ &&
            -1.444444\nh
            +1.794338\nl
            +3.238782\nm
        \biggr]
        + l_{x_{fq}}^2
        \biggl[
            68.54157
            +0.666667\nh
            \nonumber \\ &&
            -3.343491\nl
            -4.010158\nm
        \biggr]
    \biggr)
\biggr\}
+ \mathcal{O}(x_{fq}^3,\alphas^5)~,
\end{eqnarray}
where we truncate the numerical coefficients to six digits.
The definition of $x_{fq}$ can be found in Eq.~(\ref{eq::x_fq}). For the above equation
we have chosen $\mu_f^2=\mu^2=\overline{m}_1^2$.

For $x\to1$ our results for $\zmM$
and $\ymM$ read
\begin{eqnarray}
\zmM/\nm &=& \left( \frac{\alphas}{\pi} \right)^2
\biggl\{
    -0.1035672
    +0.8599121 y
    +0.177533 y^2
\biggr\}
 + \left( \frac{\alphas}{\pi} \right)^3
\biggl\{
        -0.8271467
        \nonumber \\ &&
        -0.1281609\nh
        -0.02224348\nl
        -0.06408045\nm
    + y
    \biggl(
        +13.73874
         \nonumber \\ &&
        +0.1577018\nh
        -0.4777289\nl
        +0.1577018\nm
    \biggr)
    + y^2
    \biggl(
        +5.548438
         \nonumber \\ &&
        -0.02477709\nh
        -0.2796186\nl
        -0.224962\nm
    \biggr)
\biggr\}
 + \left( \frac{\alphas}{\pi} \right)^4
\biggl\{
        +23.62352
         \nonumber \\ &&
        +1.05405\nh
        -0.07451198\nh^2
        -0.6646237\nl
        -0.07234301\nh\nl
         \nonumber \\ &&
        -0.01720209\nl^2
        +0.5270252\nm
        -0.07451198\nh\nm
        -0.03617151\nl\nm
         \nonumber \\ &&
        -0.02483733\nm^2
    + y
    \biggl(
        +252.7121
        -1.520506\nh
        +0.06793201\nh^2
         \nonumber \\ &&
        -31.30309\nl
        +0.09375969\nh\nl
        +0.6888473\nl^2
        -1.520506\nm
         \nonumber \\ &&
        +0.135864\nh\nm
        +0.09375969\nl\nm
        +0.06793201\nm^2
    \biggr)
    + y^2
    \biggl(
        +124.5568
         \nonumber \\ &&
        +0.04141809\nh
        -0.02233919\nh^2
        -13.62136\nl
        -0.04955732\nh\nl
         \nonumber \\ &&
        +0.2815718\nl^2
        -7.269009\nm
        -0.07024994\nh\nm
        +0.210506\nl\nm
         \nonumber \\ &&
        -0.04791074\nm^2
    \biggr)
\biggr\}
+ \mathcal{O}(y^3,\alphas^5)\,,
\end{eqnarray}
\begin{eqnarray}
\ymM/\nm &=& \left( \frac{\alphas}{\pi} \right)^2
\biggl\{
    +0.1035672
    -0.8599121 y_{fq}
    -0.177533 y_{fq}^2
\biggr\}
 + \left( \frac{\alphas}{\pi} \right)^3
\biggl\{
        1.859154
        \nonumber \\ &&
        +0.1281609\nh
        +0.02224348\nl
        +0.06408045\nm
    + y_{fq}
    \biggl(
        -12.59219
        \nonumber \\ &&
        -0.1577018\nh
        +0.4777289\nl
        -0.1577018\nm
    \biggr)
    + y_{fq}^2
    \biggl(
        -5.816574
        \nonumber \\ &&
        +0.02477709\nh
        +0.2796186\nl
        +0.224962\nm
    \biggr)
\biggr\}
\nonumber \\ &&
 + \left( \frac{\alphas}{\pi} \right)^4
\biggl\{
        +6.964544
        -0.4694082\nh
        +0.07451198\nh^2
        -0.9031168\nl
        \nonumber \\ &&
        +0.07234301\nh\nl
        +0.01720209\nl^2
        -0.2347041\nm
        +0.07451198\nh\nm
        \nonumber \\ &&
        +0.03617151\nl\nm
        +0.02483733\nm^2
    + y_{fq}
    \biggl(
        -222.605
        -0.4538035\nh
        \nonumber \\ &&
        -0.06793201\nh^2
        +30.45062\nl
        -0.09375969\nh\nl
        -0.6888473\nl^2
        \nonumber \\ &&
        +1.025094\nm
        -0.135864\nh\nm
        -0.09375969\nl\nm
        -0.06793201\nm^2
    \biggr)
    \nonumber \\ &&
    + y_{fq}^2
    \biggl(
        -114.9302
        -0.8370587\nh
        +0.02233919\nh^2
        +13.0182\nl
        \nonumber \\ &&
        +0.04955732\nh\nl
        -0.2815718\nl^2
        +6.416736\nm
        +0.07024994\nh\nm
        \nonumber \\ &&
        -0.210506\nl\nm
        +0.04791074\nm^2
    \biggr)
\biggr\}
+ \mathcal{O}(y_{fq}^3,\alphas^5)\,,
\end{eqnarray}
where $y=1-x$, $y_{fq}=1-x_{fq}$.
The renormalization scale has been set
as in the small-$m_2$ limit.

In the large-$m_2$ limit we have analytic results. The leading terms are given by
\begin{eqnarray}
\zmM/\nm &=& \left( \frac{\alphas}{\pi} \right)^2
\biggl\{
        -\frac{89}{432}
        +\frac{13}{18} l_z^1
        -\frac{1}{3} l_z^2
\biggr\}
 + \left( \frac{\alphas}{\pi} \right)^3
\biggl\{
                -\frac{391}{7776}
                +\frac{4 a_4}{9}
                -\frac{1327 \nh}{11664}
                \nonumber \\ &&
                -\frac{1327 \nl}{11664}
                +\frac{1685 \nm}{23328}
                -\frac{103 \pi ^4}{6480}
                -\frac{1}{54} \pi ^2 l_2^2
                +\frac{l_2^4}{54}
                +\frac{173 \zeta_3}{288}
                +\frac{2 \nh \zeta_3}{27}
                +\frac{2 \nl \zeta_3}{27}
                \nonumber \\ &&
                -\frac{7 \nm \zeta_3}{54}
        + l_z
\biggl(
                +\frac{1967}{216}
                -\frac{5 \nh}{12}
                -\frac{\nl}{12}
                +\frac{31 \nm}{162}
                +\frac{4 \pi ^2}{27}
                +\frac{2 \pi ^2 \nh}{27}
                -\frac{\pi ^2 \nl}{27}
                \nonumber \\ &&
                +\frac{2}{27} \pi ^2 l_2
                +\frac{14 \zeta_3}{9}
    \biggr)
        + l_z^2
\biggl(
                -\frac{29}{12}
                -\frac{13 \nm}{54}
    \biggr)
        + l_z^3
\biggl(
                -\frac{11}{9}
                +\frac{2 \nh}{27}
                +\frac{2 \nl}{27}
                +\frac{4 \nm}{27}
    \biggr)
\biggr\}
\nonumber \\ &&
 + \left( \frac{\alphas}{\pi} \right)^4
\biggl\{
                -\frac{19781113}{746496}
                +\frac{103045 a_4}{5184}
                +\frac{118 a_5}{27}
                +\frac{5473411 \nh}{2239488}
                -\frac{49 a_4 \nh}{108}
                \nonumber \\ &&
                -\frac{4 a_5 \nh}{9}
                -\frac{17671 \nh^2}{124416}
                +\frac{6357283 \nl}{2239488}
                -\frac{49 a_4 \nl}{108}
                -\frac{4 a_5 \nl}{9}
                -\frac{17671 \nh \nl}{62208}
                \nonumber \\ &&
                -\frac{17671 \nl^2}{124416}
                -\frac{113519017 \nm}{1097349120}
                -\frac{298241 a_4 \nm}{30240}
                +\frac{19 \nh \nm}{512}
                +\frac{19 \nl \nm}{512}
                \nonumber \\ &&
                -\frac{1281821 \nm^2}{117573120}
                +\frac{653 \pi ^2}{1944}
                +\frac{185 \pi ^2 \nh}{3888}
                -\frac{653 \pi ^2 \nl}{7776}
                +\frac{13 \pi ^2 \nm}{162}
                -\frac{2985833 \pi ^4}{7464960}
                \nonumber \\ &&
                +\frac{1549 \pi ^4 \nh}{311040}
                +\frac{7 \pi ^4 \nh^2}{8640}
                +\frac{1549 \pi ^4 \nl}{311040}
                +\frac{7 \pi ^4 \nh \nl}{4320}
                +\frac{7 \pi ^4 \nl^2}{8640}
                +\frac{970517 \pi ^4 \nm}{8709120}
                \nonumber \\ &&
                -\frac{\pi ^4 \nh \nm }{960}
                -\frac{\pi ^4 \nl \nm }{960}
                -\frac{\pi ^4 \nm^2}{25920}
                +\frac{787 \pi ^6}{81648}
                +\frac{653 \pi ^2 l_2}{3888}
                +\frac{145 \pi ^4 l_2}{1944}
                -\frac{49 \pi ^4 \nh l_2}{6480}
                \nonumber \\ &&
                -\frac{49 \pi ^4 \nl l_2}{6480}
                -\frac{103045 \pi ^2 l_2^2}{124416}
                +\frac{49 \pi ^2 \nh l_2^2}{2592}
                +\frac{49 \pi ^2 \nl l_2^2}{2592}
                +\frac{298241 \pi ^2 \nm l_2^2}{725760}
                \nonumber \\ &&
                +\frac{59}{972} \pi ^2 l_2^3
                -\frac{1}{162} \pi ^2 \nh l_2^3
                -\frac{1}{162} \pi ^2 \nl l_2^3
                +\frac{103045 l_2^4}{124416}
                -\frac{49 \nh l_2^4}{2592}
                -\frac{49 \nl l_2^4}{2592}
                \nonumber \\ &&
                -\frac{298241 \nm l_2^4}{725760}
                -\frac{59 l_2^5}{1620}
                +\frac{1}{270} \nh l_2^5
                +\frac{1}{270} \nl l_2^5
                +\frac{5029463 \zeta_3}{276480}
                +\frac{2683 \nh \zeta_3}{5184}
                \nonumber \\ &&
                +\frac{5 \nh^2 \zeta_3}{864}
                +\frac{2683 \nl \zeta_3}{5184}
                +\frac{5 \nh \nl \zeta_3}{432}
                +\frac{5 \nl^2 \zeta_3}{864}
                -\frac{85083017 \nm \zeta_3}{20321280}
                +\frac{5 \nh \nm \zeta_3}{864}
                \nonumber \\ &&
                +\frac{5 \nl \nm \zeta_3}{864}
                -\frac{17 \nm^2 \zeta_3}{1120}
                -\frac{1061 \zeta_3^2}{576}
                +\frac{62975 \zeta_5}{1728}
                -\frac{497 \nh \zeta_5}{288}
                -\frac{497 \nl \zeta_5}{288}
                -\frac{55 \nm \zeta_5}{24}
                \nonumber \\ &&
        + l_z
\biggl(
                 \frac{1186409}{6912}
                -\frac{10 a_4}{9}
                -\frac{127745 \nh}{5184}
                -\frac{4 a_4 \nh}{27}
                +\frac{2147 \nh^2}{3888}
                -\frac{72521 \nl}{5184}
                -\frac{4 a_4 \nl}{27}
                \nonumber \\ &&
                +\frac{1553 \nh \nl}{1944}
                +\frac{959 \nl^2}{3888}
                +\frac{22103 \nm}{10368}
                -\frac{4 a_4 \nm}{9}
                +\frac{143 (\nh+\nl) \nm}{1944}
                -\frac{359 \nm^2}{5832}
                \nonumber \\ &&
                +\frac{26275 \pi ^2}{2592}
                +\frac{14179 \pi ^2 \nh}{1944}
                -\frac{4}{405} \pi ^2 \nh^2
                -\frac{1021 \pi ^2 \nl}{648}
                -\frac{13}{324} \pi ^2 \nh \nl
                +\frac{13 \pi ^2 \nl^2}{324}
                \nonumber \\ &&
                -\frac{2 \pi ^2 \nm}{27}
                -\frac{3047 \pi ^4}{6480}
                +\frac{1309 \pi ^4 \nh}{19440}
                +\frac{1099 \pi ^4 \nl}{19440}
                +\frac{133 \pi ^4 \nm}{6480}
                +\frac{733}{162} \pi ^2 l_2
                \nonumber \\ &&
                -\frac{640}{81} \pi ^2 \nh l_2
                -\frac{11}{81} \pi ^2 \nl l_2
                -\frac{61}{108} \pi ^2 l_2^2
                +\frac{1}{162} \pi ^2 \nh l_2^2
                +\frac{7}{162} \pi ^2 \nl l_2^2
                +\frac{1}{54} \pi ^2 \nm l_2^2
                \nonumber \\ &&
                -\frac{5}{108} l_2^4
                -\frac{1}{162} \nh l_2^4
                -\frac{1}{162} \nl l_2^4
                -\frac{1}{54} \nm l_2^4
                +\frac{417563 \zeta_3}{13824}
                +\frac{2341 \nh \zeta_3}{864}
                -\frac{\nh^2 \zeta_3}{3}
                \nonumber \\ &&
                -\frac{3491 \nl \zeta_3}{864}
                -\frac{1}{3} \nh \nl \zeta_3
                -\frac{8269 \nm \zeta_3}{2304}
                +\frac{\nm^2 \zeta_3}{9}
                -\frac{1331 \pi ^2 \zeta_3}{432}
                -\frac{1}{4} \pi ^2 \nh \zeta_3
                \nonumber \\ &&
                -\frac{1745 \zeta_5}{216}
                +\frac{5 \nh \zeta_5}{4}
    \biggr)
        + l_z^2
\biggl(
                -\frac{54223}{1728}
                +\frac{3103 \nh}{864}
                -\frac{31 \nh^2}{324}
                +\frac{3247 \nl}{864}
                -\frac{31 \nh \nl}{162}
                \nonumber \\ &&
                -\frac{31 \nl^2}{324}
                -\frac{5995 \nm}{1728}
                +\frac{73 \nh \nm}{648}
                -\frac{35 \nl \nm}{648}
                -\frac{31 \nm^2}{324}
                +\frac{2 \pi ^2}{27}
                +\frac{\pi ^2 \nh}{27}
                -\frac{\pi ^2 \nl}{54}
                \nonumber \\ &&
                -\frac{2 \pi ^2 \nm}{27}
                -\frac{1}{27} \pi ^2 \nh \nm
                +\frac{1}{54} \pi ^2 \nl \nm
                +\frac{1}{27} \pi ^2 l_2
                -\frac{1}{27} \pi ^2 \nm l_2
                +\frac{493 \zeta_3}{36}
                \nonumber \\ &&
                -\frac{7 \nm \zeta_3}{9}
    \biggr)
        + l_z^3
\biggl(
                -\frac{995}{54}
                +\frac{391 \nh}{324}
                +\frac{391 \nl}{324}
                +\frac{176 \nm}{81}
                +\frac{13 \nm^2}{162}
    \biggr)
        + l_z^4
\biggl(
                -\frac{121}{24}
                \nonumber \\ &&
                +\frac{11 \nh}{18}
                -\frac{\nh^2}{54}
                +\frac{11 \nl}{18}
                -\frac{\nh \nl}{27}
                -\frac{\nl^2}{54}
                +\frac{35 \nm}{36}
                -\frac{\nh \nm}{18}
                -\frac{\nl \nm}{18}
                -\frac{\nm^2}{18}
    \biggr)
\biggr\}
\nonumber \\ &&
+ \mathcal{O}(z,\alphas^5)\,,
\end{eqnarray}
\begin{eqnarray}
\ymM/\nm &=& \left( \frac{\alphas}{\pi} \right)^2
\biggl\{
        +\frac{89}{432}
        -\frac{13}{18} l_{z_{fq}}
        +\frac{1}{3} l_{z_{fq}}^2
\biggr\}
 + \left( \frac{\alphas}{\pi} \right)^3
\biggl\{
                +\frac{8947}{7776}
                -\frac{4 a_4}{9}
                +\frac{1327 \nh}{11664}
                \nonumber \\ &&
                +\frac{1327 \nl}{11664}
                -\frac{1685 \nm}{23328}
                +\frac{103 \pi ^4}{6480}
                +\frac{1}{54} \pi ^2 l_2^2
                -\frac{1}{54} l_2^4
                -\frac{173 \zeta_3}{288}
                -\frac{2 \nh \zeta_3}{27}
                -\frac{2 \nl \zeta_3}{27}
                \nonumber \\ &&
                +\frac{7 \nm \zeta_3}{54}
        + l_{z_{fq}}
\biggl(
                -\frac{1759}{216}
                +\frac{5 \nh}{12}
                +\frac{\nl}{12}
                -\frac{31 \nm}{162}
                -\frac{4 \pi ^2}{27}
                -\frac{2 \pi ^2 \nh}{27}
                +\frac{\pi ^2 \nl}{27}
                \nonumber \\ &&
                -\frac{2}{27} \pi ^2 l_2
                -\frac{14 \zeta_3}{9}
    \biggr)
        + l_{z_{fq}}^2
\biggl(
                +\frac{47}{36}
                +\frac{13 \nm}{54}
    \biggr)
        + l_{z_{fq}}^3
\biggl(
                +\frac{11}{9}
                -\frac{2 \nh}{27}
                -\frac{2 \nl}{27}
                \nonumber \\ &&
                -\frac{4 \nm}{27}
    \biggr)
\biggr\}
 + \left( \frac{\alphas}{\pi} \right)^4
\biggl\{
                +\frac{36028613}{746496}
                -\frac{104581 a_4}{5184}
                -\frac{118 a_5}{27}
                -\frac{8737019 \nh}{2239488}
                \nonumber \\ &&
                +\frac{49 a_4 \nh}{108}
                +\frac{4 a_5 \nh}{9}
                +\frac{17671 \nh^2}{124416}
                -\frac{7355483 \nl}{2239488}
                +\frac{49 a_4 \nl}{108}
                +\frac{4 a_5 \nl}{9}
                \nonumber \\ &&
                +\frac{17671 \nh \nl}{62208}
                +\frac{17671 \nl^2}{124416}
                -\frac{3526303 \nm}{1097349120}
                +\frac{298241 a_4 \nm}{30240}
                -\frac{19 \nh \nm}{512}
                \nonumber \\ &&
                -\frac{19 \nl \nm}{512}
                +\frac{1281821 \nm^2}{117573120}
                +\frac{221 \pi ^2}{1944}
                +\frac{689 \pi ^2 \nh}{3888}
                -\frac{221 \pi ^2 \nl}{7776}
                -\frac{13 \pi ^2 \nm}{108}
                \nonumber \\ &&
                +\frac{3064937 \pi ^4}{7464960}
                -\frac{1549 \pi ^4 \nh}{311040}
                -\frac{7 \pi ^4 \nh^2}{8640}
                -\frac{1549 \pi ^4 \nl}{311040}
                -\frac{7 \pi ^4 \nh \nl}{4320}
                -\frac{7 \pi ^4 \nl^2}{8640}
                \nonumber \\ &&
                -\frac{970517 \pi ^4 \nm}{8709120}
                +\frac{1}{960} \pi ^4 \nh \nm
                +\frac{1}{960} \pi ^4 \nl \nm
                +\frac{\pi ^4 \nm^2}{25920}
                -\frac{787 \pi ^6}{81648}
                +\frac{221 \pi ^2 l_2}{3888}
                \nonumber \\ &&
                -\frac{145 \pi ^4 l_2}{1944}
                +\frac{49 \pi ^4 \nh l_2}{6480}
                +\frac{49 \pi ^4 \nl l_2}{6480}
                +\frac{104581 \pi ^2 l_2^2}{124416}
                -\frac{49 \pi ^2 \nh l_2^2}{2592}
                -\frac{49 \pi ^2 \nl l_2^2}{2592}
                \nonumber \\ &&
                -\frac{298241 \pi ^2 \nm l_2^2}{725760}
                -\frac{59}{972} \pi ^2 l_2^3
                +\frac{1}{162} \pi ^2 \nh l_2^3
                +\frac{1}{162} \pi ^2 \nl l_2^3
                -\frac{104581 l_2^4}{124416}
                \nonumber \\ &&
                +\frac{49 \nh l_2^4}{2592}
                +\frac{49 \nl l_2^4}{2592}
                +\frac{298241 \nm l_2^4}{725760}
                +\frac{59 l_2^5}{1620}
                -\frac{1}{270} \nh l_2^5
                -\frac{1}{270} \nl l_2^5
                \nonumber \\ &&
                -\frac{13857029 \zeta_3}{829440}
                -\frac{2939 \nh \zeta_3}{5184}
                -\frac{5 \nh^2 \zeta_3}{864}
                -\frac{2939 \nl \zeta_3}{5184}
                -\frac{5}{432} \nh \nl \zeta_3
                -\frac{5 \nl^2 \zeta_3}{864}
                \nonumber \\ &&
                +\frac{86839177 \nm \zeta_3}{20321280}
                -\frac{5(\nh+\nl) \nm \zeta_3 }{864}
                +\frac{17 \nm^2 \zeta_3}{1120}
                +\frac{1061 \zeta_3^2}{576}
                -\frac{62975 \zeta_5}{1728}
                \nonumber \\ &&
                +\frac{497 \nh \zeta_5}{288}
                +\frac{497 \nl \zeta_5}{288}
                +\frac{55 \nm \zeta_5}{24}
        + l_{z_{fq}}
\biggl(
                -\frac{1006561}{6912}
                +\frac{10 a_4}{9}
                +\frac{123565 \nh}{5184}
                \nonumber \\ &&
                +\frac{4 a_4 \nh}{27}
                -\frac{2147 \nh^2}{3888}
                +\frac{66901 \nl}{5184}
                +\frac{4 a_4 \nl}{27}
                -\frac{1553 \nh \nl}{1944}
                -\frac{959 \nl^2}{3888}
                -\frac{9925 \nm}{3456}
                \nonumber \\ &&
                +\frac{4 a_4 \nm}{9}
                -\frac{143 \nh \nm}{1944}
                -\frac{143 \nl \nm}{1944}
                +\frac{359 \nm^2}{5832}
                -\frac{985 \pi ^2}{96}
                -\frac{14299 \pi ^2 \nh}{1944}
                +\frac{4 \pi ^2 \nh^2}{405}
                \nonumber \\ &&
                +\frac{347 \pi ^2 \nl}{216}
                +\frac{13}{324} \pi ^2 \nh \nl
                -\frac{13}{324} \pi ^2 \nl^2
                +\frac{\pi ^2 \nm}{9}
                +\frac{3047 \pi ^4}{6480}
                -\frac{1309 \pi ^4 \nh}{19440}
                \nonumber \\ &&
                -\frac{1099 \pi ^4 \nl}{19440}
                -\frac{133 \pi ^4 \nm}{6480}
                -\frac{743}{162} \pi ^2 l_2
                +\frac{640}{81} \pi ^2 \nh l_2
                +\frac{11}{81} \pi ^2 \nl l_2
                +\frac{61}{108} \pi ^2 l_2^2
                \nonumber \\ &&
                -\frac{1}{162} \pi ^2 \nh l_2^2
                -\frac{7}{162} \pi ^2 \nl l_2^2
                -\frac{1}{54} \pi ^2 \nm l_2^2
                +\frac{5 l_2^4}{108}
                +\frac{1}{162} \nh l_2^4
                +\frac{1}{162} \nl l_2^4
                \nonumber \\ &&
                +\frac{1}{54} \nm l_2^4
                -\frac{128521 \zeta_3}{4608}
                -\frac{2341 \nh \zeta_3}{864}
                +\frac{\nh^2 \zeta_3}{3}
                +\frac{3491 \nl \zeta_3}{864}
                +\frac{\nh \nl \zeta_3}{3}
                \nonumber \\ &&
                +\frac{8269 \nm \zeta_3}{2304}
                -\frac{\nm^2 \zeta_3}{9}
                +\frac{1331 \pi ^2 \zeta_3}{432}
                +\frac{1}{4} \pi ^2 \nh \zeta_3
                +\frac{1745 \zeta_5}{216}
                -\frac{5 \nh \zeta_5}{4}
    \biggr)
    \nonumber \\ &&
        + l_{z_{fq}}^2
\biggl(
                +\frac{10289}{576}
                -\frac{1001 \nh}{288}
                +\frac{31 \nh^2}{324}
                -\frac{953 \nl}{288}
                +\frac{31 \nh \nl}{162}
                +\frac{31 \nl^2}{324}
                +\frac{5843 \nm}{1728}
                \nonumber \\ &&
                -\frac{73 \nh \nm}{648}
                +\frac{35 \nl \nm}{648}
                +\frac{31 \nm^2}{324}
                +\frac{2 \pi ^2}{27}
                +\frac{\pi ^2 \nh}{27}
                -\frac{\pi ^2 \nl}{54}
                +\frac{2 \pi ^2 \nm}{27}
                \nonumber \\ &&
                +\frac{1}{27} \pi ^2 \nh \nm
                -\frac{1}{54} \pi ^2 \nl \nm
                +\frac{1}{27} \pi ^2 l_2
                +\frac{1}{27} \pi ^2 \nm l_2
                -\frac{497 \zeta_3}{36}
                +\frac{7 \nm \zeta_3}{9}
    \biggr)
    \nonumber \\ &&
        + l_{z_{fq}}^3
\biggl(
                +\frac{445}{54}
                -\frac{191 \nh}{324}
                -\frac{191 \nl}{324}
                -\frac{115 \nm}{81}
                -\frac{13 \nm^2}{162}
    \biggr)
        + l_{z_{fq}}^4
\biggl(
                +\frac{121}{24}
                -\frac{11 \nh}{18}
                \nonumber \\ &&
                +\frac{\nh^2}{54}
                -\frac{11 \nl}{18}
                +\frac{\nh \nl}{27}
                +\frac{\nl^2}{54}
                -\frac{31 \nm}{36}
                +\frac{\nh \nm}{18}
                +\frac{\nl \nm}{18}
                +\frac{\nm^2}{18}
    \biggr)
\biggr\}
\nonumber \\ &&
+ \mathcal{O}({z_{fq}},\alphas^5)\,,
\end{eqnarray}
where $z=1/x$, $z_{fq}=1/x_{fq}$, $l_z=\log(z)$, $l_{z_{fq}}=\log(z_{fq})$, $l_2=\log(2)$, $a_n = \text{Li}_{n}(1/2)$.

Deeper expansions for general values of the renormalization scales for all three limiting cases
are implemented in version~3.1.3
of \texttt{RunDec}~\cite{Chetyrkin:2000yt,Schmidt:2012az,Herren:2017osy} (see also Section~\ref{sub::rundec}) which can be downloaded from~\cite{progdata_RunDec3}.
It is straightforward to extract
computer-readable expressions from
the \texttt{RunDec} source code.

\subsection{Massless limit and equal-mass case}
\label{sub::singlemass}

The limiting cases $m_2 \to 0$ and $m_2 \to m_1$ serve as an important
check on the newly calculated $\nm$ contributions to $z_m^{(4)}$. This
allows for a comparison with known results from
Refs.~\cite{Marquard:2015qpa, Marquard:2016dcn}. Indeed, by
considering the limiting cases in $\zmMfour$, it is possible to
reproduce the results for 18 of the 23 $\nm$-independent colour
structures.

\begin{table}[t]
  \centering
  {\scalefont{0.85}
  \begin{tabular}{|c|l|}
    \hline
    Colour factor & Our results  \\ \hline \hline
    $\zmFLLL = \zmFMMM(x=0)$ & $\hphantom{+}4.0688461536270908105568265845852935\textcolor{red}{9}\ldots$
    \\ \hline
    $\zmFFLL = \zmFFMM(x=0)$ & $-2.2544110367866449467998855483\textcolor{red}{0}\ldots$
    \\ \hline
    $\zmFALL = \zmFAMM(x=0)$ & $-42.4632567012766721282125905002\textcolor{red}{8}\ldots$
    \\ \hline
    $\zmFLLH = \zmFMLL(x=1)$ & $-0.103212519624620750296927052743807\textcolor{red}{7}\ldots$
    \\ \hline
    $\zmFHHH = \zmFMMM(x=1)$ & $-0.14902396167117774491351258459995232994034582577\textcolor{red}{8}\ldots$
    \\ \hline
    $\zmFLHH = \zmFMML(x=1)$ & $-0.21702903967506766885962266353048096302320696185\textcolor{red}{2}\ldots$
    \\ \hline
    $\zmFFLH = \zmFFML(x=1)$ & $-2.892087945496480288157357839486299242092348766\textcolor{red}{6}\ldots$
    \\ \hline
    $\zmFALH = \zmFAML(x=1)$ & $\hphantom{+}0.6207486755507333054358038056222730447120129137\textcolor{red}{8}\ldots$
    \\ \hline
  \end{tabular}
  }
  \caption{Comparison of the numerical and analytical results (see Refs.~\cite{Lee:2013sx,Fael:2021kyg}), which can be obtained by setting $x=0$ or $x=1$ in
  the respective terms containing two heavy quark flavours.
  The red digit indicates the first deviation
  from the analytical result.
  }
  \label{tab::zm4_exact}
\end{table}

In Tab.~\ref{tab::zm4_exact}, we list the eight four-loop coefficients
for which analytical results were obtained in Refs.~\cite{Lee:2013sx,Fael:2021kyg}. For
illustration, we compare these with the numerical results obtained
using the ``expand-and-match'' method. We present the numerical
results up to the point where a discrepancy first appears relative to
the numerical evaluation of the analytical expressions. An impressive
agreement of 29 or more significant digits is observed. It should be
noted that as input for the ``expand-and-match'' procedure we use the
exact analytical result for $x\to\infty$ and transport it to $x=1$ or
$x=0$, respectively. The results for $x=1$ are generally even
more precise than those for $x=0$.

\begin{table}[t]
  \centering
{\scalefont{0.85}
  \begin{tabular}{|c|rcl|l|}
    \hline
    Colour factor & \multicolumn{3}{c}{Results of Ref.~\cite{Marquard:2016dcn}} & Our results  \\ \hline \hline
    $\zmdFFL = \zmdFFM(x=0)$ & $\textcolor{Green}{-0.9}37 $&$\pm$&$ 0.178$ & $-0.95044828732178848273...$ \\ \hline
    $\zmdFFH = \zmdFFM(x=1)$ & $\textcolor{Green}{-3.9}24 $&$\pm$&$    0.642$ &  $-3.9475522724259017481...$ \\ \hline
    $\zmFFFL = \zmFFFM(x=0)$ & $\textcolor{Green}{-0.05}094 $&$\pm$&$    0.00298$ &  $-0.051273375289534988607...$ \\ \hline
    $\zmFFAL = \zmFFAM(x=0)$ & $\textcolor{Green}{9.266}42 $&$\pm$&$    0.00454$ &  $\hphantom{+}9.2669169668474129435...$ \\ \hline
    $\zmFAAL = \zmFAAM(x=0)$ & $\textcolor{Green}{122.187}2 $&$\pm$&$    0.0100$ &  $\hphantom{+}122.18776459581115274...$ \\ \hline
    $\zmFFFH = \zmFFFM(x=1)$ & $\textcolor{Green}{-1.36}25 $&$\pm$&$    0.0132$ &  $-1.3646701555565992063...$ \\ \hline
    $\zmFFAH = \zmFFAM(x=1)$ & $\textcolor{Green}{14.98}00 $&$\pm$&$    0.0334$ &  $\hphantom{+}14.984457292343298328...$ \\ \hline
    $\zmFAAH = \zmFAAM(x=1)$ & $\textcolor{Green}{-2.3}597 $&$\pm$&$    0.0342$ &  $-2.3615622565916925413...$ \\ \hline
    $\zmFFHH = \zmFFMM(x=1)$ & $\textcolor{Green}{1.6575}2 $&$\pm$&$    0.00031$ & $\hphantom{+}1.6575134347128087589...$ \\ \hline
    $\zmFAHH = \zmFAMM(x=1)$ & $\textcolor{Green}{-0.209}34 $&$\pm$&$    0.00273$ &  $-0.20964741309703695536...$ \\  \hline
  \end{tabular}
}
\caption{Comparison of our results with the ones of Ref.~\cite{Marquard:2016dcn}, which can be obtained by setting $x=0$ or $x=1$ in
  the respective terms containing two heavy quark flavours.
  The green digits indicate digits which agree between both calculations.
  We find good agreement often even one digit beyond the accuracy stated in Ref.~\cite{Marquard:2016dcn}.
  }
  \label{tab::zm4_old_new}
\end{table}

The results for the remaining 10 coefficients, for which no analytical
results are available, are listed in Tab.~\ref{tab::zm4_old_new}. In
the second column, we show the results from Ref.~\cite{Marquard:2016dcn}
and highlight in green those digits that match the numerical results
obtained in the present work (presented in the third column). In many
cases where the method used in the earlier calculation yielded only a
few significant digits, we are able to significantly increase the
precision. In the third column, we provide 20 significant digits for
each colour structure. Based on considerations regarding the
cancellation of $1/\epsilon$ poles (see Section~\ref{sec::ren}) and
the results discussed in Tab.~\ref{tab::zm4_exact}, it can be inferred
that our precision is even higher; however, to remain on the safe
side, we quote a precision of 20 significant digits.

\subsection{Numerical effects}

In this subsection, we briefly discuss the numerical effect of a
second finite quark mass for the cases $m_1=m_b$ and $m_1=m_t$. For
the numerical evaluation, we use $\alphas^{(5)}(m_Z)=0.118$ and treat
$m_1^\mathrm{OS} = \overline{m}_1(\mu_1) y_m$ as an expansion in
$\alphas^{(\nf)}(\mu_1)$, with $\nf=6$ for $m_1=m_t$ and
$\nf=5$ for $m_1=m_b$.
$y_m$ furthermore depends on $x_{fq}(\mu_f,\mu)$, see Eq.~(\ref{eq::x_fq}).
We choose
$\overline{m}_t(\overline{m}_t)=163$~GeV,
$\overline{m}_b(\overline{m}_b)=4.2$~GeV, and
$\overline{m}_c(3~\mbox{GeV})=0.986$~GeV as input values for the quark
masses. For the computation of
$\alphas^{(6)}(\mu)$ we use the
decoupling relation expressed in terms of the pole mass for the heavy quark. We use $m_t^{\rm OS}=172$~GeV
and perform the decoupling at the scale $\mu_{\rm dec}=2m_t^{\rm OS}$.
We use \texttt{RunDec}~\cite{Chetyrkin:2000yt,Schmidt:2012az,Herren:2017osy} to perform the running and decoupling to the relevant order in $\alphas$.

For the bottom quark mass we
obtain for $\mu_1=\overline{m}_b(\overline{m}_b)$
and $\alphas^{(5)}$
the following perturbative expansion
\begin{eqnarray}
   \frac{m_b^{\rm OS}}{\mbox{MeV}} &=& 4\,200 + \left(400\right)_{\alphas}
   + \left(199 + 7\right)_{\alphas^2}
   + \left( 144 + 16\right)_{\alphas^3}
   + \left( 135 + 29\right)_{\alphas^4}
   + {\cal O}(\alphas^5)
   \nonumber\\
   \label{eq::mbnum}
\end{eqnarray}
where the subscripts indicate the order in
$\alphas$. In each round bracket the first number
corresponds to the
vanishing-mass approximation and the second to the contribution of a
finite charm-quark mass. In the limit of a massless charm quark, the
second number would be zero.

For the bottom-quark-mass effects to the top-quark pole mass
we have
\begin{eqnarray}
   \frac{m_t^{\rm OS}}{\mbox{MeV}} &=&
   163\,000
   + \left( 7\,504\right)_{\alphas}
   + \left( 1\,601 + 8\right)_{\alphas^2}
   + \left( 494 + 11\right)_{\alphas^3}
   + \left( 194 + 13\right)_{\alphas^4}
   + {\cal O}(\alphas^5)
   \,,
   \nonumber\\
   \label{eq::mtnum}
\end{eqnarray}
where $\mu_1=\overline{m}_t(\overline{m}_t)$ and $\alphas^{(6)}$ is used
as expansion parameter.

The typical loop momentum at order $\alphas^{n+1}$ is of the order of
magnitude $m_Q\re^{-n}$~\cite{Ball:1995ni,Beneke:2016cbu}.
Therefore,
it is to be expected that finite quark masses become increasingly
important at higher loop orders
which is in agreement with the numerical results in
Eqs.~(\ref{eq::mbnum}) and~(\ref{eq::mtnum}).
It has been proposed in the literature (see, e.g.,
Ref.~\cite{Ayala:2014yxa})
to decouple the heavy quarks from the running of $\alphas$ and to
consider the expansion in terms of $\alphas^{(3)}$ -- for both the
bottom and the top quark.
Following Ref.~\cite{Ayala:2014yxa} we obtain for the bottom quark mass relation
\begin{eqnarray}
   \frac{m_b^{\rm OS}}{\mbox{MeV}} &\!=& 4\,200 + \left(382\right)_{\alphas}
   \!+ \left(202.0 - 0.5\right)_{\alphas^2}
   \!+ \left(155.7 - 0.2\right)_{\alphas^3}
   \!+ \left(155.4 - 0.2\right)_{\alphas^4}
   \!+ {\cal O}(\alphas^5).
   \nonumber\\
   \label{eq::mbnum_as3}
\end{eqnarray}
As expected, we now observe that the mass correction terms are small. Up
to order $\alphas^3$ our results are compatible\footnote{It should be noted that the cancellation of the original mass correction
terms and those arising from the transition from $\alphas^{(5)}$
to $\alphas^{(3)}$ is quite subtle. Consequently, the final numerical values
are highly sensitive to the specific values of the input parameters
for $\alphas$, $m_c$ and $m_b$.} with those from Ref.~\cite{Ayala:2014yxa}. The four-loop terms are new.

It should be noted that the analytical expression leading to
Eq.~(\ref{eq::mbnum_as3}) contains logarithmic terms of the type
$\log(m_c/m_b)$. Since it is numerically small, no resummation
is required. The situation is different when effects of finite charm
and bottom masses are investigated in the context of mass relations
for the top quark. In this case, the resummation of logarithms such as
$\log(m_c/m_t)$ and $\log(m_b/m_t)$ is mandatory. This can be achieved
using, for example, the formalism developed in Ref.~\cite{Hoang:2017btd}.

\subsection{Comparison to predictions of the four-loop light-mass correction terms}

\begin{figure}[t]
    \centering
    \includegraphics[width=0.95\linewidth]{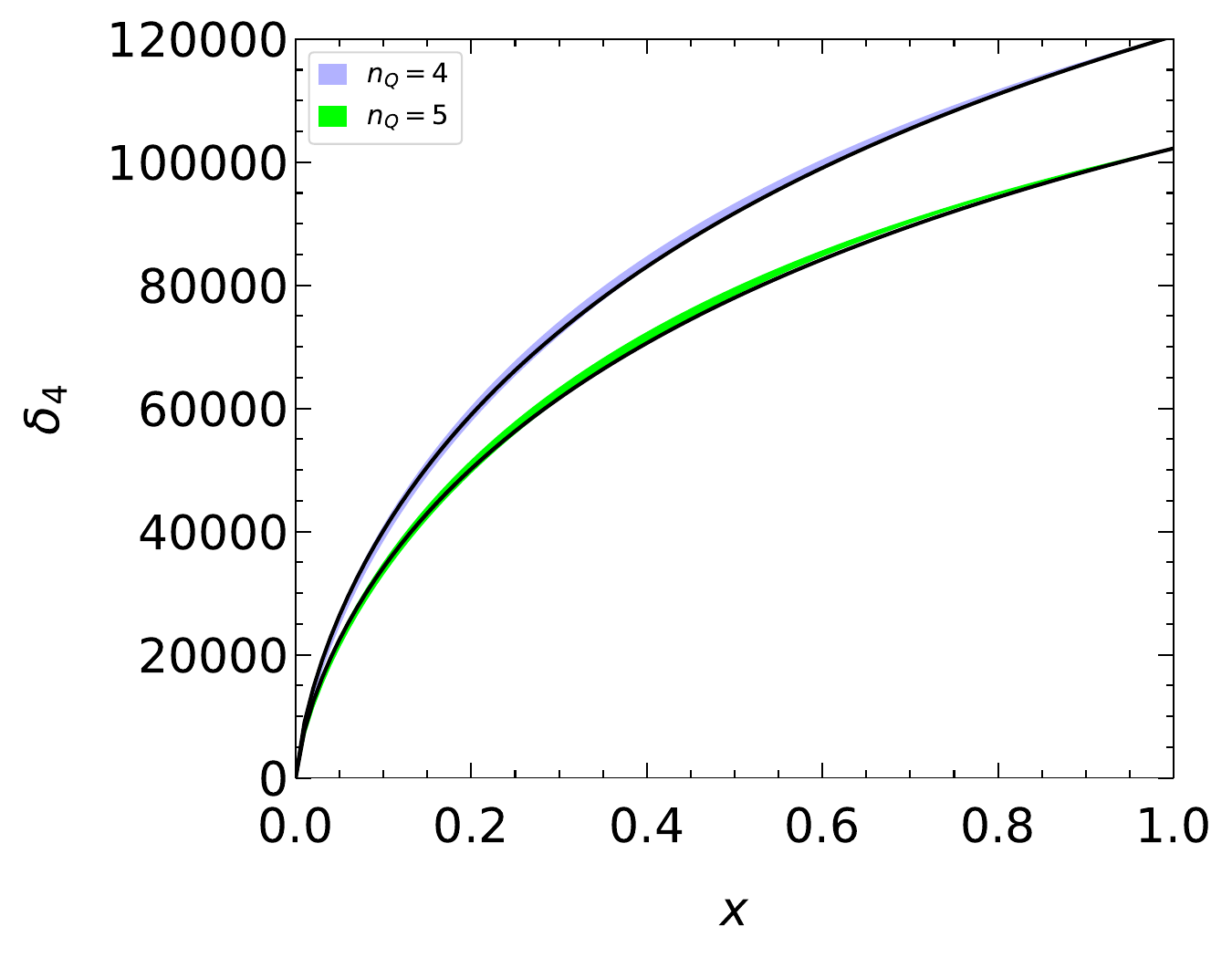}
    \caption{Comparison of the exact results (black lines) and the approximations, see text for details.}
    \label{fig::compare}
\end{figure}

\begin{figure}[t]
    \centering
    \includegraphics[width=0.48\linewidth]{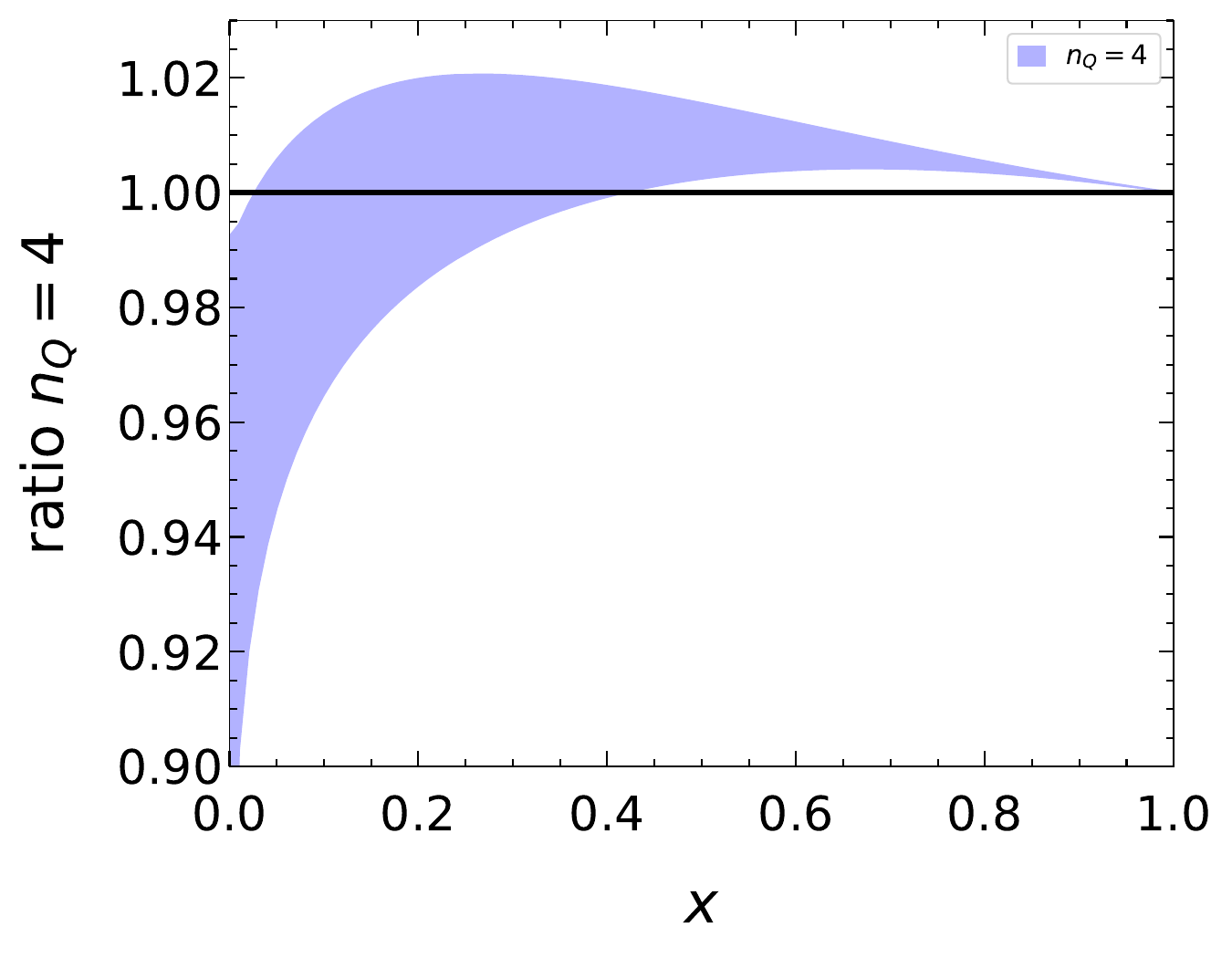}
    \includegraphics[width=0.48\linewidth]{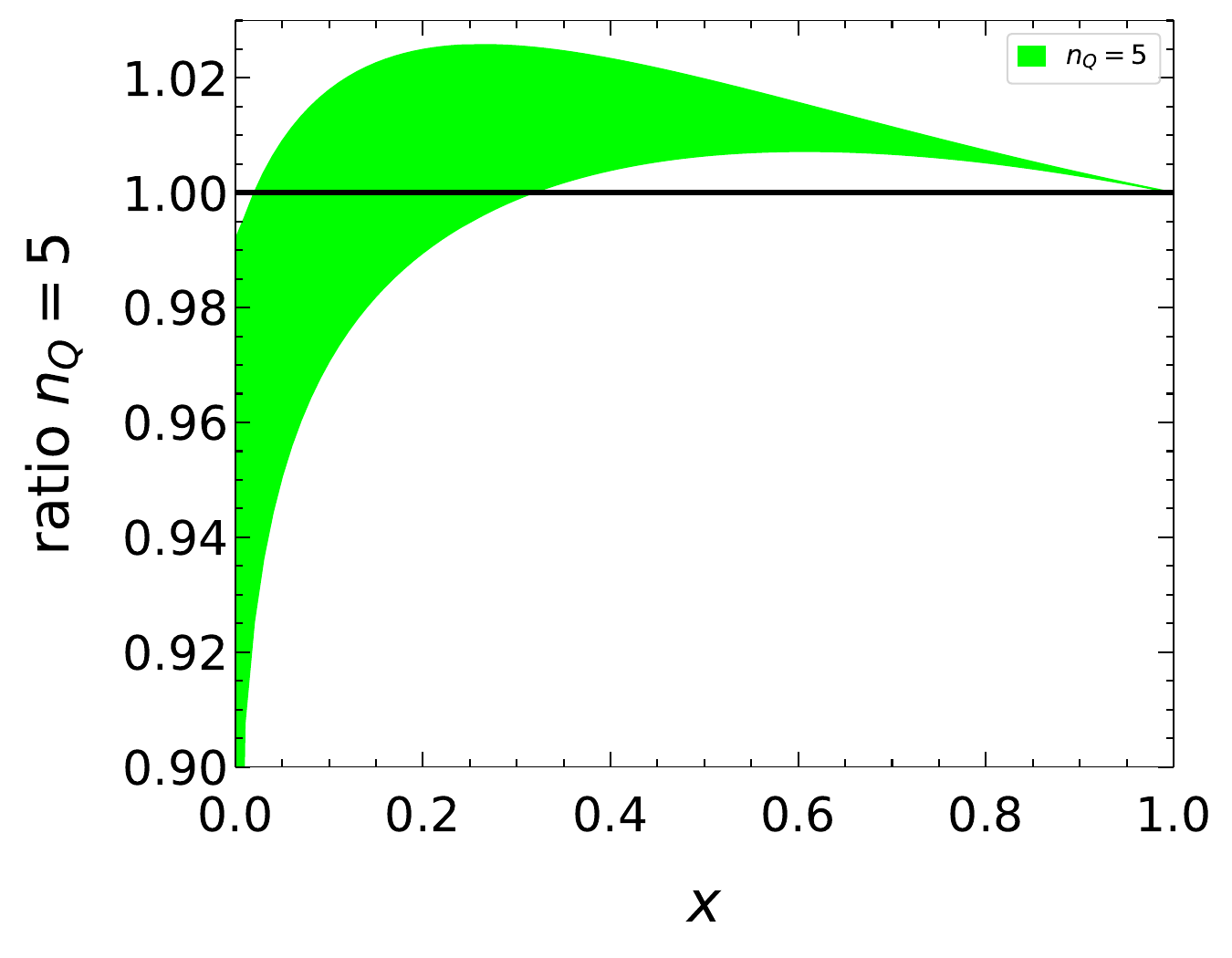}
    \caption{Ratio of the curves shown in Fig.~\ref{fig::compare} to the exact result.
    }
    \label{fig::compare_ratio}
\end{figure}

A subset of the corrections induced by a second massive quark (as
computed in the present paper) has been predicted in
Refs.~\cite{Hoang:2017btd,Mateu:2017hlz} based on heavy-quark symmetry
and the cancellation of renormalon contributions.  In the following we
compare our results with the approximation given in Eq.~(3.12) of
Ref.~\cite{Hoang:2017btd}
(see also Eq.~(3.14) of that paper), which
agree within uncertainties with the predictions in
Ref.~\cite{Mateu:2017hlz}.

Before making the comparison, we must process our result as follows:
\begin{itemize}
\item Consider all contributions with $\nm^n$ with $n=1,2,3$.
\item Set $\nh=0$ to exclude contributions from closed
   loops from quarks with mass $m_1$.
\item Introduce the renormalization scale for the mass $m_2$,
 i.e., express the result for $\ymM$ in terms of $\alphas(\mu)$,
 $m_1(\mu)$ and $m_2(\mu_f)$ (see also Eq.~(\ref{eq::x_fq})).
\item Set $\mu=m_1(\mu)$ and $\mu_f=m_2(\mu_f)$.
\item The strong coupling constant is still defined in the $\nl+\nm+\nh$ scheme.
\item Consider the quantity $\ymM(x_{fq}) - \ymM(x_{fq}=0)$
  which vanishes for $m_2=0$.
\item We denote the four-loop corrections of the resulting quantity by $\delta_4$, which is shown in Fig.~\ref{fig::compare}.
In Eq.~(3.12) of Ref.~\cite{Hoang:2017btd} this
quantity is denoted by $\delta^{(q)}_{Q,4}$.
\end{itemize}

In Fig.~\ref{fig::compare}, we show the exact results (constructed as
described in the previous paragraph) as well as the approximations for
$n_Q \equiv \nl + 1 = 4$ and $n_Q = 5$; these correspond to the
bottom-quark self-energy with a massive charm quark and the top-quark
self-energy with a massive bottom quark, respectively. The
approximations are also displayed in the left panel of Fig.~4 of
Ref.~\cite{Hoang:2017btd}, which we reproduce in
Fig.~\ref{fig::compare}. The band arises from varying the internal
scale, which is present in the approximation formula Eq.~(3.12) of
Ref.~\cite{Hoang:2017btd}, between $m_2(m_2)$ and $m_1(m_1)$ (see the
caption for Fig.~4 in \cite{Hoang:2017btd}). For greater clarity,
Fig.~\ref{fig::compare_ratio}
shows the ratios relative to the exact
result for $n_Q = 4$ and $n_Q = 5$.

Overall, we observe good agreement between the predictions and the
exact results. It should be noted that for $x$-values relevant to the
bottom-charm system ($x \approx 0.3$), the uncertainty band of the
approximation is approximately 2\% and encompasses the exact
result. For the top-bottom system, where $x \approx 0.02$, the
uncertainty band has a width of about 8\% and lies slightly below the
exact result. By construction, the bands shrink to zero as $x \to 0$
and $x \to 1$. As a consequence the relative error shown in Fig.~\ref{fig::compare_ratio} becomes large.
It should be noted that for $x \gtrsim 0.4$ (with
$n_Q=4$) or $x \gtrsim 0.3$ (with $n_Q=5$), the exact result lies
below the uncertainty band.
The $\nh$ contribution, which is not included in the approximation discussed above,
accounts for approximately $0.1-0.2\%$ of the total result
at $x=0.02$ and $x=0.3$
and is therefore very small.

\subsection{Implementation in {\rm RunDec} and {\rm CRunDec}}
\label{sub::rundec}

The new corrections calculated in this work have been implemented in
\texttt{RunDec}~\cite{Chetyrkin:2000yt,Schmidt:2012az,Herren:2017osy}.\footnote{See also \texttt{REvolver}~\cite{Hoang:2021fhn}, a program with similar functionality.} The following functions have been
updated:\footnote{We refer to the publications on
\texttt{RunDec}~\cite{Chetyrkin:2000yt,Schmidt:2012az,Herren:2017osy} for a description of the functions and
the meaning of the arguments.}
\begin{verbatim}
mMS2mOS[mMS, {mq}, asmu, mu, nf, l]
mOS2mMS[mOS, {mq}, asmu, mu, nf, l]
mOS2mSI[mOS, {mq}, asM, nf, l]
\end{verbatim}
The parameter \verb|nf| represents the total number of active
quarks.
Accounting for the mass effects of a second massive quark
is controlled via the argument \verb|{mq}|,
with the following options available:
\begin{itemize}

\item \verb|{}| (empty set): All secondary quarks are massless; only $m_1$ is different from zero.

\item \verb|{m2}|: Here \verb|m2| is interpreted as $\overline{m}_2(\overline{m}_2)$.
  The corrections for the secondary mass depend on
  $\overline{m}_2(\overline{m}_2)/\overline{m}_1(\mu)$
  in the case of \verb|mMS2mOS|
  and on $\overline{m}_2(\overline{m}_2)/m_1^{\rm OS}$
  in the case of   \verb|mOS2mMS|
  and \verb|mOS2mSI|.

\item \verb|{{m2,muf}}|:
  Here \verb|muf|=$\mu_f$ is the renormalization scale of $\overline{m}_2$.
  The corrections for the secondary mass depend on
  $\overline{m}_2(\mu_f)/\overline{m}_1(\mu)$
  in the case of \verb|mMS2mOS|
  and on $\overline{m}_2(\mu_f)/m_1^{\rm OS}$, in the case of
  \verb|mOS2mMS| and \verb|mOS2mSI|.

\item \verb|{m2,m3,...}| and \verb|{{m2,muf2},{m3,muf3},...}|:
  This notation is used to include the corrections
  from several secondary massive quarks.

  It should be noted that results from Feynman diagrams involving
  three different massive quarks are not yet available and are
  therefore not included in \texttt{RunDec}. Such corrections appear
  for the first time at three-loop order.

\end{itemize}
The corrections for the additional masses are implemented such that,
as $m_2, m_3, \ldots \to 0$, the results for \verb|{}| are
recovered. There is no restriction regarding the ratio $m_2/m_1$. In
particular, $m_2/m_1 > 1$ is also permissible.

The latest version of \texttt{RunDec} can be downloaded
from~\cite{progdata_RunDec3}. It contains concrete examples illustrating the
options discussed above. In particular, it demonstrates how the
numerical results from Eqs.~(\ref{eq::mbnum}) and~(\ref{eq::mtnum})
are obtained.

Analogous functions are implemented in the \texttt{C++} version of
\texttt{RunDec}: \texttt{CRunDec}. We again provide examples which
illustrate the use of the new options.

\section{\label{sec::concl}Conclusions and outlook}

In this work, we presented the determination of the relation
between the $\overline{\rm MS}$ and on-shell quark masses up to
four-loop order, accounting for two non-vanishing quark mass
scales. Using standard techniques, we reduce the underlying Feynman
integrals to a small set of master integrals. The
corresponding differential equation with respect to the ratio of the
two mass scales, $x=m_2/m_1$, can be solved using a semi-analytical
approach, ``expand and match'', which yields expansions around suitably
chosen values of $x$. A major advantage of this approach
is that the differential equation does not need to be in a specific
form. The boundary conditions can be calculated analytically for the
limiting case $x\to\infty$.

We have determined the mass renormalization constants with high
precision to ten or more significant digits across the entire range of
phenomenologically relevant mass ratios. In the appropriate limiting
cases, our results reproduce the known expressions for the
single-scale case, providing a rigorous consistency check of the
calculation. Furthermore, we significantly improve the precision for
various colour structures.

The numerical evaluation of the conversion relation between the
$\overline{\rm MS}$ and on-shell mass shows that the light-quark mass effects become larger at higher loop order.
In particular, for the bottom quark the effects induced by the
finite charm-quark mass amount to
about 7~MeV at two loops and 29~MeV
at four loops when the perturbative series is parametrized in terms of
$\alphas$ with five active quark flavours. If instead $\alphas^{(3)}$ is used, the light-quark mass effects
are small and amount to only about $0.2$~MeV at four loops.

We compare our results after switching off contributions from closed heavy quark loops with predictions of the light-quark mass effects available in the literature. We find agreement at the few percent level.

Interesting next steps include calculating similar corrections for the
relation between the on-shell mass and so-called short-distance
masses (kinetic mass, potential-subtracted mass, 1S mass, etc.),
enabling the establishment of a precise relation between the
$\overline{\rm MS}$ mass and the short-distance masses
accounting for corrections due to finite light-quark masses.
In particular, our results provide the four-loop finite light-quark mass corrections needed to extend the relation between the pole and MSR masses to this order.
Another obvious application is the calculation of similar corrections for the
on-shell wave-function renormalization constant. In this case,
however, it is necessary to reduce integrals featuring an additional
``dot'', which makes the reduction to master
integrals an even greater challenge.


\section*{Acknowledgements}

The work of F.L.~was supported by the Swiss National Science Foundation (SNSF) under contract \href{https://data.snf.ch/grants/grant/211209}{TMSGI2\_211209}.
This work was supported by the Deutsche Forschungsgemeinschaft (DFG, German
Research Foundation) under grant 396021762 --- TRR 257 ``Particle Physics
Phenomenology after the Higgs Discovery''.
K.S.~is supported by the European Union under the Marie Sk{\l}odowska-Curie Actions (MSCA) Grant 101204018.
We thank Andr\'e Hoang for useful discussions in connection
to Ref.~\cite{Hoang:2017btd}.
We thank Florian Herren for implementing the results obtained in this work into \texttt{CRunDec}.
We thank Andr\'e Hoang and Vicent Mateu
for useful comments on the manuscript.

\bibliographystyle{jhep}
\bibliography{MSOSm1m2.bib,extra.bib}

@article{book_smirnov_2012,
author = {Smirnov, Vladimir},
year = {2012},
month = {01},
pages = {},
title = {Analytic Tools for Feynman Integrals},
volume = {250},
isbn = {978-3-642-34885-3},
journal = {Springer Tracts Mod. Phys.},
doi = {10.1007/978-3-642-34886-0}
}

@article{Mateu:2017hlz,
    author = "Mateu, Vicent and Ortega, Pablo G.",
    title = "{Bottom and charm mass determinations from global fits
    $Q\bar{Q}$ bound states at N$^3$LO}",
    eprint = "1711.05755",
    archivePrefix = "arXiv",
    primaryClass = "hep-ph",
    doi = "10.1007/JHEP01(2018)122",
    journal = "JHEP",
    volume = "01",
    pages = "122",
    year = "2018"
}

@article{Chetyrkin:2015mxa,
    author = {Chetyrkin, Konstantin G. and K{\"u}hn, Johann H. and Steinhauser, Matthias and Sturm, Christian},
    editor = {Bl{\"u}mlein, Johannes and Jansen, Karl and Kr{\"a}mer, Michael and K{\"u}hn, Johann H.},
    title = "{Massive Tadpoles: Techniques and Applications}",
    eprint = "1502.00509",
    archivePrefix = "arXiv",
    primaryClass = "hep-ph",
    reportNumber = "TTP14-039, SFB-CPP-14-104",
    doi = "10.1016/j.nuclphysbps.2015.03.003",
    journal = "Nucl. Part. Phys. Proc.",
    volume = "261-262",
    pages = "19--30",
    year = "2015"
}

@article{Schroder:2005va,
    author = "Schröder, Y. and Vuorinen, A.",
    title = "{High-precision epsilon expansions of single-mass-scale four-loop vacuum bubbles}",
    eprint = "hep-ph/0503209",
    archivePrefix = "arXiv",
    reportNumber = "BI-TP-2005-11, UW-PT-05-5",
    doi = "10.1088/1126-6708/2005/06/051",
    journal = "JHEP",
    volume = "06",
    pages = "051",
    year = "2005"
}

@article{Lee:2010ik,
    author = "Lee, R. N. and Smirnov, V. A.",
    title = "{Analytic epsilon expansion of three-loop on-shell master integrals up to four-loop transcendentality weight}",
    eprint = "1010.1334",
    archivePrefix = "arXiv",
    primaryClass = "hep-ph",
    doi = "10.1007/JHEP02(2011)102",
    journal = "JHEP",
    volume = "02",
    pages = "102",
    year = "2011"
}

@article{Ayala:2014yxa,
    author = "Ayala, C{\'e}sar and Cveti{\v{c}}, Gorazd and Pineda, Antonio",
    title = "{The bottom quark mass from the $ \boldsymbol{\Upsilon} (1S) $ system at NNNLO}",
    eprint = "1407.2128",
    archivePrefix = "arXiv",
    primaryClass = "hep-ph",
    reportNumber = "USM-TH-326",
    doi = "10.1007/JHEP09(2014)045",
    journal = "JHEP",
    volume = "09",
    pages = "045",
    year = "2014"
}

@article{Herren:2017osy,
    author = "Herren, Florian and Steinhauser, Matthias",
    title = "{Version 3 of RunDec and CRunDec}",
    eprint = "1703.03751",
    archivePrefix = "arXiv",
    primaryClass = "hep-ph",
    reportNumber = "TTP17-011",
    doi = "10.1016/j.cpc.2017.11.014",
    journal = "Comput. Phys. Commun.",
    volume = "224",
    pages = "333--345",
    year = "2018"
}

@article{Schmidt:2012az,
    author = "Schmidt, Barbara and Steinhauser, Matthias",
    title = "{CRunDec: A C++ package for running and decoupling of the strong coupling and quark masses}",
    eprint = "1201.6149",
    archivePrefix = "arXiv",
    primaryClass = "hep-ph",
    reportNumber = "SFB-CPP-12-03, TTP12-02",
    doi = "10.1016/j.cpc.2012.03.023",
    journal = "Comput. Phys. Commun.",
    volume = "183",
    pages = "1845--1848",
    year = "2012"
}

@article{Chetyrkin:2000yt,
    author = "Chetyrkin, K. G. and Kühn, Johann H. and Steinhauser, M.",
    title = "{RunDec: a Mathematica package for running and decoupling of the strong coupling and quark masses}",
    eprint = "hep-ph/0004189",
    archivePrefix = "arXiv",
    reportNumber = "DESY-00-034, TTP-00-05",
    doi = "10.1016/S0010-4655(00)00155-7",
    journal = "Comput. Phys. Commun.",
    volume = "133",
    pages = "43--65",
    year = "2000"
}

@article{Hoang:2021fhn,
    author = "Hoang, Andre H. and Lepenik, Christopher and Mateu, Vicent",
    title = "{REvolver: Automated running and matching of couplings and masses in QCD}",
    eprint = "2102.01085",
    archivePrefix = "arXiv",
    primaryClass = "hep-ph",
    reportNumber = "UWThPh-2020-18, IFT-UAM/CSIC-21-5",
    doi = "10.1016/j.cpc.2021.108145",
    journal = "Comput. Phys. Commun.",
    volume = "270",
    pages = "108145",
    year = "2022"
}

@article{Lee:2013sx,
    author = "Lee, Roman and Marquard, Peter and Smirnov, Alexander V. and Smirnov, Vladimir A. and Steinhauser, Matthias",
    title = "{Four-loop corrections with two closed fermion loops to fermion self energies and the lepton anomalous magnetic moment}",
    eprint = "1301.6481",
    archivePrefix = "arXiv",
    primaryClass = "hep-ph",
    reportNumber = "SFB-CPP-13-03, TTP13-02",
    doi = "10.1007/JHEP03(2013)162",
    journal = "JHEP",
    volume = "03",
    pages = "162",
    year = "2013"
}

@article{Fael:2020bgs,
    author = {Fael, Matteo and Sch{\"o}nwald, Kay and Steinhauser, Matthias},
    title = "{Exact results for $ {Z}_m^{\mathrm{OS}} $ and $ {Z}_2^{\mathrm{OS}} $ with two mass scales and up to three loops}",
    eprint = "2008.01102",
    archivePrefix = "arXiv",
    primaryClass = "hep-ph",
    reportNumber = "TTP20-028, P3H-20-038",
    doi = "10.1007/JHEP10(2020)087",
    journal = "JHEP",
    volume = "10",
    pages = "087",
    year = "2020"
}

@article{Chetyrkin:1999ys,
    author = "Chetyrkin, K. G. and Steinhauser, M.",
    title = "{Short-Distance Mass of a Heavy Quark at Order $\alpha_s^3$}",
    eprint = "hep-ph/9907509",
    archivePrefix = "arXiv",
    reportNumber = "BUTP-99-15, TTP-99-33",
    doi = "10.1103/PhysRevLett.83.4001",
    journal = "Phys. Rev. Lett.",
    volume = "83",
    pages = "4001--4004",
    year = "1999"
}

@article{Chetyrkin:1999qi,
    author = "Chetyrkin, K. G. and Steinhauser, M.",
    title = "{The relation between the $\overline{MS}$ and the on-shell quark mass at order $\alpha_s^3$}",
    eprint = "hep-ph/9911434",
    archivePrefix = "arXiv",
    reportNumber = "DESY-99-174, TTP-99-47",
    doi = "10.1016/S0550-3213(99)00784-1",
    journal = "Nucl. Phys. B",
    volume = "573",
    pages = "617--651",
    year = "2000"
}

@article{Ball:1995ni,
    author = "Ball, Patricia and Beneke, M. and Braun, Vladimir M.",
    title = "{Resummation of $(\beta_0 \alpha_s)^n$ corrections in QCD: techniques and applications to the $\tau$ hadronic width and the heavy quark pole mass}",
    eprint = "hep-ph/9502300",
    archivePrefix = "arXiv",
    reportNumber = "CERN-TH-95-26, CERN-TH-95-026, UM-TH-95-3",
    doi = "10.1016/0550-3213(95)00392-6",
    journal = "Nucl. Phys. B",
    volume = "452",
    pages = "563--625",
    year = "1995"
}

@article{Peraro:2016wsq,
    author = "Peraro, Tiziano",
    title = "{Scattering amplitudes over finite fields and multivariate functional reconstruction}",
    eprint = "1608.01902",
    archivePrefix = "arXiv",
    primaryClass = "hep-ph",
    reportNumber = "EDINBURGH-2016-14",
    doi = "10.1007/JHEP12(2016)030",
    journal = "JHEP",
    volume = "12",
    pages = "030",
    year = "2016"
}

@article{Klappert:2020aqs,
    author = "Klappert, Jonas and Klein, Sven Yannick and Lange, Fabian",
    title = "{Interpolation of dense and sparse rational functions and other improvements in FireFly}",
    eprint = "2004.01463",
    archivePrefix = "arXiv",
    primaryClass = "cs.MS",
    reportNumber = "TTK-20-07, P3H-20-010",
    doi = "10.1016/j.cpc.2021.107968",
    journal = "Comput. Phys. Commun.",
    volume = "264",
    pages = "107968",
    year = "2021"
}

@article{Liu:2021wks,
    author = "Liu, Xiao and Ma, Yan-Qing",
    title = "{Multiloop corrections for collider processes using auxiliary mass flow}",
    eprint = "2107.01864",
    archivePrefix = "arXiv",
    primaryClass = "hep-ph",
    doi = "10.1103/PhysRevD.105.L051503",
    journal = "Phys. Rev. D",
    volume = "105",
    number = "5",
    pages = "L051503",
    year = "2022"
}

@article{Gerlach:2022qnc,
    author = "Gerlach, Marvin and Herren, Florian and Lang, Martin",
    title = "{tapir: A tool for topologies, amplitudes, partial fraction decomposition and input for reductions}",
    eprint = "2201.05618",
    archivePrefix = "arXiv",
    primaryClass = "hep-ph",
    reportNumber = "TTP22-005, P3H-22-009, FERMILAB-PUB-22-025-T",
    doi = "10.1016/j.cpc.2022.108544",
    journal = "Comput. Phys. Commun.",
    volume = "282",
    pages = "108544",
    year = "2023"
}

@article{Lee:2013mka,
    author = "Lee, Roman N.",
    editor = "Wang, Jianxiong",
    title = "{LiteRed 1.4: a powerful tool for reduction of multiloop integrals}",
    eprint = "1310.1145",
    archivePrefix = "arXiv",
    primaryClass = "hep-ph",
    doi = "10.1088/1742-6596/523/1/012059",
    journal = "J. Phys. Conf. Ser.",
    volume = "523",
    pages = "012059",
    year = "2014"
}

@article{vonManteuffel:2014ixa,
    author = "von Manteuffel, Andreas and Schabinger, Robert M.",
    title = "{A novel approach to integration by parts reduction}",
    eprint = "1406.4513",
    archivePrefix = "arXiv",
    primaryClass = "hep-ph",
    reportNumber = "MITP-14-009",
    doi = "10.1016/j.physletb.2015.03.029",
    journal = "Phys. Lett. B",
    volume = "744",
    pages = "101--104",
    year = "2015"
}

@article{Remiddi:1997ny,
    author = "Remiddi, Ettore",
    title = "{Differential equations for Feynman graph amplitudes}",
    eprint = "hep-th/9711188",
    archivePrefix = "arXiv",
    reportNumber = "DFUB-97-15, DFUB 97-15",
    doi = "10.1007/BF03185566",
    journal = "Nuovo Cim. A",
    volume = "110",
    pages = "1435--1452",
    year = "1997"
}

@article{Liu:2022chg,
    author = "Liu, Xiao and Ma, Yan-Qing",
    title = "{AMFlow: A Mathematica package for Feynman integrals computation via auxiliary mass flow}",
    eprint = "2201.11669",
    archivePrefix = "arXiv",
    primaryClass = "hep-ph",
    doi = "10.1016/j.cpc.2022.108565",
    journal = "Comput. Phys. Commun.",
    volume = "283",
    pages = "108565",
    year = "2023"
}

@article{Liu:2017jxz,
    author = "Liu, Xiao and Ma, Yan-Qing and Wang, Chen-Yu",
    title = "{A systematic and efficient method to compute multi-loop master integrals}",
    eprint = "1711.09572",
    archivePrefix = "arXiv",
    primaryClass = "hep-ph",
    doi = "10.1016/j.physletb.2018.02.026",
    journal = "Phys. Lett. B",
    volume = "779",
    pages = "353--357",
    year = "2018"
}

@article{Usovitsch:2020jrk,
    author = "Usovitsch, Johann",
    title = "{Factorization of denominators in integration-by-parts reductions}",
    eprint = "2002.08173",
    archivePrefix = "arXiv",
    primaryClass = "hep-ph",
    month = "2",
    year = "2020"
}

@article{Chetyrkin:2004mf,
    author = "Chetyrkin, K. G.",
    title = "{Four-loop renormalization of QCD: full set of renormalization constants and anomalous dimensions}",
    eprint = "hep-ph/0405193",
    archivePrefix = "arXiv",
    reportNumber = "TTP04-08, SFB-CPP-04-16",
    doi = "10.1016/j.nuclphysb.2005.01.011",
    journal = "Nucl. Phys. B",
    volume = "710",
    pages = "499--510",
    year = "2005"
}

@article{Klappert:2020nbg,
    author = {Klappert, Jonas and Lange, Fabian and Maierh{\"o}fer, Philipp and Usovitsch, Johann},
    title = "{Integral reduction with Kira 2.0 and finite field methods}",
    eprint = "2008.06494",
    archivePrefix = "arXiv",
    primaryClass = "hep-ph",
    reportNumber = "TTK-20-24, P3H-20-041, FR-PHENO-2020-11, MITP/20-044",
    doi = "10.1016/j.cpc.2021.108024",
    journal = "Comput. Phys. Commun.",
    volume = "266",
    pages = "108024",
    year = "2021"
}

@article{Klappert:2019emp,
    author = "Klappert, Jonas and Lange, Fabian",
    title = "{Reconstructing rational functions with FireFly}",
    eprint = "1904.00009",
    archivePrefix = "arXiv",
    primaryClass = "cs.SC",
    reportNumber = "TTK-19-11, P3H-19-007",
    doi = "10.1016/j.cpc.2019.106951",
    journal = "Comput. Phys. Commun.",
    volume = "247",
    pages = "106951",
    year = "2020"
}

@article{Lee:2012cn,
    author = "Lee, R. N.",
    title = "{Presenting LiteRed: a tool for the Loop InTEgrals REDuction}",
    eprint = "1212.2685",
    archivePrefix = "arXiv",
    primaryClass = "hep-ph",
    month = "12",
    year = "2012"
}

@article{Gray:1990yh,
    author = "Gray, N. and Broadhurst, David J. and Grafe, W. and Schilcher, K.",
    title = "{Three-loop relation of $\overline{MS}$ quark and pole masses}",
    reportNumber = "OUT-4102-28",
    doi = "10.1007/BF01614703",
    journal = "Z. Phys. C",
    volume = "48",
    pages = "673--680",
    year = "1990"
}

@article{Smirnov:2020quc,
    author = "Smirnov, A. V. and Smirnov, V. A.",
    title = "{How to choose master integrals}",
    eprint = "2002.08042",
    archivePrefix = "arXiv",
    primaryClass = "hep-ph",
    reportNumber = "P3H-20-008, TTP20-008",
    doi = "10.1016/j.nuclphysb.2020.115213",
    journal = "Nucl. Phys. B",
    volume = "960",
    pages = "115213",
    year = "2020"
}

@article{Marquard:2015qpa,
    author = "Marquard, Peter and Smirnov, Alexander V. and Smirnov, Vladimir A. and Steinhauser, Matthias",
    title = "{Quark Mass Relations to Four-Loop Order in Perturbative QCD}",
    eprint = "1502.01030",
    archivePrefix = "arXiv",
    primaryClass = "hep-ph",
    reportNumber = "DESY-15-013, TTP15-01, SFB-CPP-14-120",
    doi = "10.1103/PhysRevLett.114.142002",
    journal = "Phys. Rev. Lett.",
    volume = "114",
    number = "14",
    pages = "142002",
    year = "2015"
}

@article{Tarrach:1980up,
    author = "Tarrach, R.",
    title = "{The pole mass in perturbative QCD}",
    reportNumber = "CPT-80/P-1215",
    doi = "10.1016/0550-3213(81)90140-1",
    journal = "Nucl. Phys. B",
    volume = "183",
    pages = "384--396",
    year = "1981"
}

@article{Marquard:2007uj,
    author = "Marquard, P. and Mihaila, L. and Piclum, J. H. and Steinhauser, M.",
    title = "{Relation between the pole and the minimally subtracted mass in dimensional regularization and dimensional reduction to three-loop order}",
    eprint = "hep-ph/0702185",
    archivePrefix = "arXiv",
    reportNumber = "DESY-07-016, SFB-CPP-07-04, TTP-07-04",
    doi = "10.1016/j.nuclphysb.2007.03.010",
    journal = "Nucl. Phys. B",
    volume = "773",
    pages = "1--18",
    year = "2007"
}

@article{Liu:2022mfb,
    author = "Liu, Zhi-Feng and Ma, Yan-Qing",
    title = "{Determining Feynman Integrals with Only Input from Linear Algebra}",
    eprint = "2201.11637",
    archivePrefix = "arXiv",
    primaryClass = "hep-ph",
    doi = "10.1103/PhysRevLett.129.222001",
    journal = "Phys. Rev. Lett.",
    volume = "129",
    number = "22",
    pages = "222001",
    year = "2022"
}

@article{Hoang:2017btd,
    author = "Hoang, Andre H. and Lepenik, Christopher and Preisser, Moritz",
    title = "{On the light massive flavor dependence of the large order asymptotic behavior and the ambiguity of the pole mass}",
    eprint = "1706.08526",
    archivePrefix = "arXiv",
    primaryClass = "hep-ph",
    reportNumber = "UWTHPH-2017-13, MIT-CTP-4915",
    doi = "10.1007/JHEP09(2017)099",
    journal = "JHEP",
    volume = "09",
    pages = "099",
    year = "2017"
}

@article{Beneke:2016cbu,
    author = "Beneke, M. and Marquard, P. and Nason, P. and Steinhauser, M.",
    title = "{On the ultimate uncertainty of the top quark pole mass}",
    eprint = "1605.03609",
    archivePrefix = "arXiv",
    primaryClass = "hep-ph",
    reportNumber = "DESY-16-082, TTP16-016, TUM-HEP-1044-16",
    doi = "10.1016/j.physletb.2017.10.054",
    journal = "Phys. Lett. B",
    volume = "775",
    pages = "63--70",
    year = "2017"
}

@article{Gehrmann:1999as,
    author = "Gehrmann, T. and Remiddi, E.",
    title = "{Differential equations for two-loop four-point functions}",
    eprint = "hep-ph/9912329",
    archivePrefix = "arXiv",
    reportNumber = "TTP-99-49",
    doi = "10.1016/S0550-3213(00)00223-6",
    journal = "Nucl. Phys. B",
    volume = "580",
    pages = "485--518",
    year = "2000"
}

@article{Nogueira:1991ex,
    author = "Nogueira, Paulo",
    title = "{Automatic Feynman Graph Generation}",
    reportNumber = "IFM-7-91",
    doi = "10.1006/jcph.1993.1074",
    journal = "J. Comput. Phys.",
    volume = "105",
    pages = "279--289",
    year = "1993"
}

@article{Fael:2021kyg,
    author = {Fael, Matteo and Lange, Fabian and Sch{\"o}nwald, Kay and Steinhauser, Matthias},
    title = "{A semi-analytic method to compute Feynman integrals applied to four-loop corrections to the $ \overline{\mathrm{MS}} $-pole quark mass relation}",
    eprint = "2106.05296",
    archivePrefix = "arXiv",
    primaryClass = "hep-ph",
    reportNumber = "P3H-21-041, TTP21-016",
    doi = "10.1007/JHEP09(2021)152",
    journal = "JHEP",
    volume = "09",
    pages = "152",
    year = "2021"
}

@article{Tkachov:1981wb,
    author = "Tkachov, F. V.",
    title = "{A theorem on analytical calculability of 4-loop renormalization group functions}",
    doi = "10.1016/0370-2693(81)90288-4",
    journal = "Phys. Lett. B",
    volume = "100",
    pages = "65--68",
    year = "1981"
}

@article{Fael:2021xdp,
    author = {Fael, Matteo and Lange, Fabian and Sch{\"o}nwald, Kay and Steinhauser, Matthias},
    title = "{A semi-numerical method for one-scale problems applied to the $\overline{\rm MS}$-on-shell relation}",
    eprint = "2110.03699",
    archivePrefix = "arXiv",
    primaryClass = "hep-ph",
    reportNumber = "TTP21-033, P3H-21-067",
    doi = "10.21468/SciPostPhysProc.7.041",
    journal = "SciPost Phys. Proc.",
    volume = "7",
    pages = "041",
    year = "2022"
}

@article{Lee:2015eva,
    author = "Lee, Roman N. and Mingulov, Kirill T.",
    title = "{Introducing SummerTime: A package for high-precision computation of sums appearing in DRA method}",
    eprint = "1507.04256",
    archivePrefix = "arXiv",
    primaryClass = "hep-ph",
    doi = "10.1016/j.cpc.2016.02.018",
    journal = "Comput. Phys. Commun.",
    volume = "203",
    pages = "255--267",
    year = "2016"
}

@article{Davies:2026cci,
    author = "Davies, J. and Kaneko, T. and Marinissen, C. and Ueda, T. and Vermaseren, J. A. M.",
    title = "{FORM Version 5.0}",
    eprint = "2601.19982",
    archivePrefix = "arXiv",
    primaryClass = "hep-ph",
    month = "1",
    year = "2026"
}

@inproceedings{Seidensticker:1999bb,
    author = "Seidensticker, T.",
    title = "{Automatic application of successive asymptotic expansions of Feynman diagrams}",
    booktitle = "{6th International Workshop on New Computing Techniques in Physics Research: Software Engineering, Artificial Intelligence Neural Nets, Genetic Algorithms, Symbolic Algebra, Automatic Calculation}",
    eprint = "hep-ph/9905298",
    archivePrefix = "arXiv",
    reportNumber = "TTP-99-22",
    month = "5",
    year = "1999"
}

@article{Fael:2022miw,
    author = {Fael, Matteo and Lange, Fabian and Sch{\"o}nwald, Kay and Steinhauser, Matthias},
    title = "{Singlet and nonsinglet three-loop massive form factors}",
    eprint = "2207.00027",
    archivePrefix = "arXiv",
    primaryClass = "hep-ph",
    reportNumber = "TTP22-042, P3H-22-066",
    doi = "10.1103/PhysRevD.106.034029",
    journal = "Phys. Rev. D",
    volume = "106",
    number = "3",
    pages = "034029",
    year = "2022"
}

@article{Chetyrkin:1981qh,
    author = "Chetyrkin, K. G. and Tkachov, F. V.",
    title = "{Integration by parts: The algorithm to calculate $\beta$-functions in 4 loops}",
    doi = "10.1016/0550-3213(81)90199-1",
    journal = "Nucl. Phys. B",
    volume = "192",
    pages = "159--204",
    year = "1981"
}

@article{Maierhofer:2017gsa,
    author = {Maierh{\"o}fer, Philipp and Usovitsch, Johann and Uwer, Peter},
    title = "{Kira{\textemdash}A Feynman integral reduction program}",
    eprint = "1705.05610",
    archivePrefix = "arXiv",
    primaryClass = "hep-ph",
    doi = "10.1016/j.cpc.2018.04.012",
    journal = "Comput. Phys. Commun.",
    volume = "230",
    pages = "99--112",
    year = "2018"
}

@article{Smirnov:2023yhb,
    author = "Smirnov, Alexander V. and Zeng, Mao",
    title = "{FIRE 6.5: Feynman integral reduction with new simplification library}",
    eprint = "2311.02370",
    archivePrefix = "arXiv",
    primaryClass = "hep-ph",
    doi = "10.1016/j.cpc.2024.109261",
    journal = "Comput. Phys. Commun.",
    volume = "302",
    pages = "109261",
    year = "2024"
}

@article{Kauers:2008zz,
    author = "Kauers, Manuel",
    editor = "Blumlein, J. and Moch, S. and Riemann, T.",
    title = "{Fast Solvers for Dense Linear Systems}",
    doi = "10.1016/j.nuclphysbps.2008.09.111",
    journal = "Nucl. Phys. B Proc. Suppl.",
    volume = "183",
    pages = "245--250",
    year = "2008"
}

@article{Melnikov:2000qh,
    author = "Melnikov, Kirill and van Ritbergen, Timo",
    title = "{The three-loop relation between the $\overline{MS}$ and the pole quark masses}",
    eprint = "hep-ph/9912391",
    archivePrefix = "arXiv",
    reportNumber = "SLAC-PUB-8321, TTP-99-51",
    doi = "10.1016/S0370-2693(00)00507-4",
    journal = "Phys. Lett. B",
    volume = "482",
    pages = "99--108",
    year = "2000"
}

@article{Bekavac:2007tk,
    author = "Bekavac, S. and Seidel, D. and Steinhauser, M. and Grozin, A. ",
    title = "{Light quark mass effects in the on-shell renormalization constants}",
    eprint = "0708.1729",
    archivePrefix = "arXiv",
    primaryClass = "hep-ph",
    reportNumber = "SFB-CPP-07-47, TTP07-21",
    doi = "10.1088/1126-6708/2007/10/006",
    journal = "JHEP",
    volume = "10",
    pages = "006",
    year = "2007"
}

@article{Laporta:2000dsw,
    author = "Laporta, S.",
    title = "{High-precision calculation of multiloop Feynman integrals by difference equations}",
    eprint = "hep-ph/0102033",
    archivePrefix = "arXiv",
    doi = "10.1142/S0217751X00002159",
    journal = "Int. J. Mod. Phys. A",
    volume = "15",
    pages = "5087--5159",
    year = "2000"
}

@article{Harlander:1998cmq,
    author = "Harlander, R. and Seidensticker, T. and Steinhauser, M.",
    title = "{Complete corrections of $O(\alpha \alpha_s)$ to the decay of the Z boson into bottom quarks}",
    eprint = "hep-ph/9712228",
    archivePrefix = "arXiv",
    reportNumber = "MPI-PHT-97-81, TTP-97-52",
    doi = "10.1016/S0370-2693(98)00220-2",
    journal = "Phys. Lett. B",
    volume = "426",
    pages = "125--132",
    year = "1998"
}

@article{Kotikov:1990kg,
    author = "Kotikov, A. V.",
    title = "{Differential equations method. New technique for massive Feynman diagram calculation}",
    reportNumber = "ITF-90-31E",
    doi = "10.1016/0370-2693(91)90413-K",
    journal = "Phys. Lett. B",
    volume = "254",
    pages = "158--164",
    year = "1991"
}

@article{Marquard:2016dcn,
    author = "Marquard, Peter and Smirnov, Alexander V. and Smirnov, Vladimir A. and Steinhauser, Matthias and Wellmann, David",
    title = "{$\overline{\rm MS}$-on-shell quark mass relation up to four loops in QCD and a general SU$(N)$ gauge group}",
    eprint = "1606.06754",
    archivePrefix = "arXiv",
    primaryClass = "hep-ph",
    reportNumber = "DESY-16-106, TTP16-023",
    doi = "10.1103/PhysRevD.94.074025",
    journal = "Phys. Rev. D",
    volume = "94",
    number = "7",
    pages = "074025",
    year = "2016"
}

@misc{progdata_RunDec3,
  author = "",
  date = "",
  howpublished = "RunDec 3.1.3 can be downloaded from: \url{https://www.ttp.kit.edu/preprints/2017/ttp17-011}"
}

@misc{exp_git,
  author = "",
  date = "",
  howpublished = "\url{https://gitlab.com/m-steinhauser/exp}"
}

@misc{fermat,
      author         = "Lewis, R. H.",
      title          = "{\it Computer Algebra System {\tt Fermat}}",
      howpublished            = "\url{https://home.bway.net/lewis/}"
}

@article{pslq,
 ISSN = {00255718, 10886842},
 URL = {http://www.jstor.org/stable/2585116},
 author = {Helaman R. P. Ferguson and David H. Bailey and Steve Arno},
 journal = {Mathematics of Computation},
 number = {225},
 pages = {351--369},
 publisher = {American Mathematical Society},
 title = {Analysis of PSLQ, An Integer Relation Finding Algorithm},
 urldate = {2024-03-05},
 volume = {68},
 year = {1999}
}

\end{document}